\documentclass[useAMS,usenatbib]{mn2e}

\usepackage{graphicx}
\usepackage{txfonts}

\newcommand{\mgtwo}{Mg$_{2}$}

\title[Stellar abundance gradients]{Stellar abundance gradients in galactic
disks. I. Method and spectral line gradients}
   \author[S. D. Ryder, Y. Fenner and B. K. Gibson]
   {S. D.  Ryder$^{1}$\thanks{E-mail: sdr@aaoepp.aao.gov.au, yfenner@swin.edu.au}, 
   Y. Fenner$^{2}$\footnotemark[1]
   and  B. K. Gibson$^{2}$\\
   $^{1}$Anglo-Australian Observatory, P.O. Box 296, Epping, NSW 1710,
   Australia\\
   $^{2}$Centre for Astrophysics and Supercomputing, Swinburne University of
   Technology, Hawthorn VIC 3122, Australia}

\begin{document}

\date{}

\pagerange{\pageref{firstpage}--\pageref{lastpage}} \pubyear{2004}

\maketitle

\label{firstpage}

\begin{abstract}    
We describe the technique of absorption line imaging of galaxy disks
using the Taurus Tunable Filter on the Anglo-Australian Telescope and
demonstrate its sensitivity to the behaviour of spectral features
associated with Mg and Fe.  Radial profiles of {\mgtwo} and Fe5270
line-strengths are presented for a sample of eight face-on spiral
galaxies spanning a range of Hubble types. Signatures of phenomena
including merger-induced star formation, \mbox{H\,{\sc ii}} rings and
galactic bars are also reported. This study demonstrates the capacity
of tunable filters to measure Mg and Fe line-strengths across the face
of spiral galaxies, which can ultimately reveal clues about the star
formation history and chemical evolution.
\end{abstract} 

      

\section{Introduction}

Radial gradients in the abundances of elements such as oxygen,
nitrogen, and sulphur have long been observed in the disk of the Milky
Way \citep{Sha83}, as well as in the disks of many nearby spiral
galaxies \citep{ZKH94, Ryd95}. The gas-phase abundances of these
elements in \mbox{H\,{\sc ii}}~regions are relatively straightforward
to determine on the basis of their bright emission-line ratios (see
the review by \citet{HW99}). Analyses by \citet{VCE92} and others have
revealed interesting correlations between the slope of these abundance
gradients, and global galaxy characteristics.  A variety of mechanisms
have been proposed to account for the fact that the inner regions of
spiral galaxies are generally more metal-rich than the outer parts,
including ongoing infall of metal-poor gas, and radial inflow of
enriched gas. The presence of a bar has been shown to homogenise the
abundances somewhat, resulting in a flatter abundance profile than in
a non-barred galaxy of similar Hubble type \citep{Mar98}.
 
Since oxygen is a primary element produced in massive stars, its
abundance is a useful measure of the {\em cumulative\/} massive star
formation rate. In order to build up a proper star formation {\em
history\/} however, it is necessary to be able to measure the relative
abundances of other elements such as iron and magnesium, which are
produced in different types of stars and released into the
interstellar medium on different characteristic timescales.  Abundance
determinations from absorption line studies of these elements are
commonplace for stars in our own Galaxy (e.g. \citealt{FJ93};
\citealt{GWJ95}); in the integrated light of elliptical galaxies,
bulges, and extragalactic globular clusters (e.g. Huchra et al. 1996;
Trager et al. 1998; Cohen, Blakeslee, \& C{\^ o}t{\' e} 2003); and as
a function of radius for inner galactic regions (Raimann et al. 2001;
Jablonka, Gorgas, \& Goudfrooij 2002; Proctor \& Sansom 2002; Worthey
2004). However, such measurements become much more difficult in
galactic disks, due to the fact that the surface brightness of the
unresolved stellar population (against which a weak absorption feature
must be measured) is much lower. Fortunately, interference filter
imaging offers almost an {\em order of magnitude\/} gain in effective
signal-to-noise over a long-slit spectrograph of the same resolution,
as demonstrated by \citet{BH97} and Moll{\' a}, Hardy \& Beauchamp
(1999). Despite a modest telescope aperture, and the need for
expensive, custom-made filters with broader bandpasses than one would
like ($\sim60$\,\AA, to be able to cover a useful range in galaxy
redshift), these pioneering studies have revealed the existence of
shallow radial gradients in the Mg$_{2}
\lambda 5176$ and Fe$\lambda 5270$ features in the disks of NGC~4303,
NGC~4321, and NGC~4535.

In this paper, we describe the application of narrow-band
absorption-line imaging techniques using the wavelength agility, wide
field-of-view, and charge-shuffling ability of the Taurus Tunable
Filter (TTF) at the Anglo-Australian Telescope (AAT) to almost triple
the number of face-on spiral galaxies for which stellar radial
abundance gradients have been measured. We begin by outlining the
TTF's characteristics in Section~\ref{method}, before demonstrating
the sensitivity of our technique on stars with a range of known
abundances in Section~\ref{stellarobs}. Our galaxy imaging and surface
photometry techniques are presented in Sections~\ref{galobs} and
\ref{datareductionanalysis}, and the radial gradients in ``Lick/IDS''
spectral line indices (Faber et al. 1985; Worthey et al. 1994) in
Section~\ref{results}. The Lick spectral indices do not measure
abundances {\em per se}, but can be transformed to quantities like
[Fe/H] via the method of spectral synthesis. In Paper~II (Fenner,
Gibson \& Ryder, in preparation), we present models and tools for
converting between absolute elemental abundances and line indices in
the observer's plane, and thus place new constraints on the star
formation histories of the galaxies presented here.


\section{Absorption Line Imaging}\label{method}

\subsection{The Taurus Tunable Filter (TTF) Instrument}

The TTF is a tunable Fabry-Perot Interferometer consisting of two
highly polished glass plates whose separation is controlled to high
accuracy by piezoelectric stacks. Unlike most Fabry-Perot etalons
which have resolving powers $R>1000$, the TTF is designed to work at
plate separations $\sim10~\mu$m or less, delivering $R=100 - 1000$,
comparable to conventional fixed narrow-band interference filters (see
\citet{BHJ98} and \citet{BHKC03} for summaries of the optical design
and practical applications of the TTF). The TTF is mounted in the
collimated beam of the Taurus~II focal reducer, and at the f/8
Cassegrain focus of the 3.9~m AAT delivers a field of view up to
10~arcmin in diameter (depending on the clear aperture of the blocking
filter in use). The EEV $2048\times$4096 CCD detector used for these
observations has a scale of 0.33~arcsec~pixel$^{-1}$, and a quantum
efficiency near 90~percent in the $5000-6000$\,\AA\ wavelength range.

By enabling an adjustable passband anywhere between 3700 and
10000\,\AA, the TTF obviates the need for an entire suite of
narrow-band filters, and makes possible monochromatic imaging of
almost any feature of interest over an expansive redshift range.
Another important aspect of our observing method is the use of
charge-shuffling synchronised with passband-switching, which overcomes
many of the systematic errors which plague conventional narrow-band
imaging, while allowing accurate differential measurements under less
than ideal observing conditions.

\subsection{Choice of Lick indices and blocking filters}\label{blockingfilters}

In their work, \citet{BH97} and \citet{MHB99} concentrated on the
{\mgtwo} and Fe5270 features, two of the most prominent ``Lick/IDS''
indices (\citealt{FFBG85}; \citealt{WFGB94}), in part to limit the
number of expensive narrow-band filters that would be required. We
chose to image our sample galaxies in these same two lines in order to
be consistent with their study and to facilitate comparison with the
large body of observational and theoretical work performed with these
two indices. Nevertheless, as \citet{TB95} point out, these features
are susceptible to some contamination by other elements, such as Ca
and C.

To simplify the transformations from our observed measurements of the
absorption line and continuum flux ratios to the Lick indices, we have
endeavoured to match as closely as possible the original line and
continuum bandpass definitions laid down by \citet{FFBG85}. While
selecting the narrowest possible bandpass for the TTF will ensure the
maximum contrast between the absorption line core flux and continuum
band, the overall signal-to-noise (S/N) of each measurement (and more
significantly, of their ratio which defines the index) will suffer
from the reduced throughput.  On the other hand, the broadest bandpass
will improve the throughput, but at the expense of diluting the
absorption feature. Because narrowing the gap between the plates in
order to broaden the passband risks irreparable damage to the plates
should their coatings come into contact with each other, we opted for
a minimum plate spacing of 2.5~microns, which yields a Lorentzian
passband with Full Width at Half-Maximum (FWHM) $\sim15$\,\AA\ at
5100\,\AA. This is four times narrower than the fixed filters employed
by \citet{BH97}, and being of the same order as the equivalent widths
of the {\mgtwo} and Fe5270 features, is a good compromise between
contrast and S/N.

Like all Fabry-Perot etalons, the TTF has a periodic transmission
profile, with a finesse $N=40$, i.e. an inter-order spacing of
$40\times$ the instrumental FWHM. Thus, to ensure only a single order
of the TTF reaches the detector, we require a blocking filter with
${\rm FWHM}<600$\,\AA, but broad enough to contain the full {\mgtwo}
and Fe5270 passbands. Unfortunately, none of the existing TTF blocking
filters meets both these requirements, and the f126 blocking filter we
used, a 380\,\AA\ FWHM interference filter centred on 5220\,\AA, was not
ideal. As Fig.~\ref{stellarscan_fig} shows, the blue continuum of
the {\mgtwo} feature lies just outside the blue cutoff of the f126
filter, while the transmission peaks at barely 70~percent, and varies
by up to 35~percent over just 100\,\AA. In addition, the clear aperture
of just 63~mm limits the usable field of view to about 6~arcmin in
diameter. Nevertheless, as we show in the following sections, this
blocking filter adequately met the goals of this project. To cover the
blue continuum of the {\mgtwo} feature, we used the TTF's B4 blocking
filter, with a FWHM = 320\,\AA\ centred on 5000\,\AA, a clear aperture
of 122~mm, and a peak throughput of 76~percent.

\subsection{Wavelength calibration}\label{wave_calib}

The plate separation and parallelism in the TTF is set by three
piezoelectric stacks, and servo-stabilised against drifting by a
capacitance bridge. Tuning to the desired wavelength is performed by
fine adjustment of the plate separation through a Queensgate
Instruments CS100 controller. In order to define the relationship
between a given plate separation ($Z$, in analog-to-digital units) and
central wavelength of the TTF passband ($\lambda$, in \AA), we
illuminate the TTF uniformly with an arc lamp, and record the lamp
spectrum while scanning the TTF progressively in wavelength.  For the
region of interest ($4800-6400$\,\AA), a combination of deuterium,
helium, and neon lamps yielded sufficient lines for this purpose.

As noted by \citet{JSBH02}, at plate separations $\lesssim 3$~microns
the multi-layer coating thickness becomes a significant fraction of
the actual plate spacing. Since the depth within the coatings where
reflection occurs is wavelength-dependent, the net effect is that the
relation between $Z$ and $\lambda$ is no longer linear as it is for
larger plate spacings (or equivalently, narrower bandpasses).  Indeed,
we found it necessary to apply a quadratic (second order) fit to the
$Z(\lambda)$ relation. Incorrectly assuming a linear relation could
result in the passband being offset by the equivalent of the FWHM at
the extreme wavelength settings.


\section{Stellar observations and calibration onto the Lick system}\label{stellarobs}

In order to derive a function for converting absorption line-strengths
measured with the TTF into the Lick/IDS system, we applied our
technique to 31 Lick standard stars for which {\mgtwo} and Fe5270
indices are available in Worthey et~al. (1994). These stars spanned a
broad range in metallicity, and thus in their Lick indices. Stars were
imaged at $\sim$6$\,\AA$ intervals in a wavelength range from a
minimum of either 5100$\,\AA$ or 4870\,$\AA$ to a maximum 5400\,$\AA$,
with the TTF set to a bandpass of FWHM $\sim$17\,$\AA$. In this way, a
low-resolution spectrum was built-up from the series of individual
images for each star. Each stellar image was trimmed, bias corrected,
and flux within a fixed aperture measured using tasks within the {\sc
iraf} package.

\begin{figure}
\includegraphics[width=8.7cm]{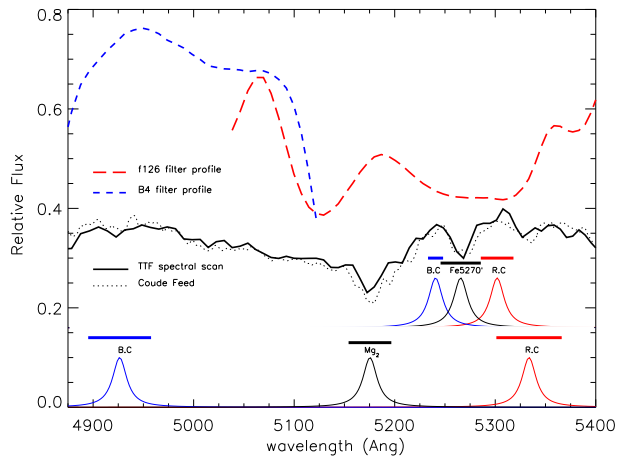}
      \caption{TTF spectral scan (\emph{solid curve}) for the star
      HD131977 versus a Coud\'{e} Feed stellar spectrum, degraded to
      match the resolution of the TTF data (\emph{dotted curve}). The
      TTF spectrum has been corrected for the transmission profiles of
      the f126 and B4 blocking filters (\emph{dashed lines}). The
      Lorentzian profiles with FWHM of 17\,$\AA$ illustrate the shape
      of the TTF bandpass. They are centred on the {\mgtwo} and Fe5270
      absorption features and their corresponding red and blue
      continua, whose locations were defined by \citealt{FFBG85}.
      Thick horizontal lines indicate the Lick bandpasses.}
      \label{stellarscan_fig}
\end{figure}

Figure~\ref{stellarscan_fig} compares a TTF spectral scan (\emph{solid
line}) for the star HD131977 against a Coud\'{e} Feed stellar spectrum
of the same star (found at \emph{The Indo-U.S. Library of Coud\'{e}
Feed Stellar Spectra} home page at http://www.noao.edu/cflib/),
degraded to match the resolution of the TTF data (\emph{dotted
line}). The TTF spectrum has been corrected for the transmission
profiles of the f126 and B4 blocking filters (described in
Section~\ref{blockingfilters}). The filter transmission profiles
(\emph{dashed lines}) were obtained by dividing our TTF spectral scans
of the flux standards HR7596 and HR5501 by the flux calibrated spectra
from \citet{Ham92}.  The alignment between the spectral features seen
in the TTF scan and those of the Coud\'{e} Feed spectrum confirms that
the quadratic $Z(\lambda)$ relation described in
Section~\ref{wave_calib} leads to good wavelength calibration. The
location of the Lorentzian profiles shown in
Fig.~\ref{stellarscan_fig} indicates the central wavelengths of the
{\mgtwo} and Fe5270 absorption features and the corresponding red and
blue continua, as defined by \citet{FFBG85}. The shape of the
Lorentzians illustrates that of the TTF bandpass used in this study,
particularly the broad wings.

The full stellar spectral scans provide a means for checking not just
the wavelength calibration, but just as importantly, in gauging the
sensitivity of this imaging technique to spectral features. However,
the {\mgtwo} and Fe5270 indices were not calculated from the full
spectral scans. Instead, they were derived using only the images at
the (velocity-corrected) line and continuum wavelengths, since this is
the technique to be applied to the galaxies.  Instrumental indices
were calculated from the following expressions ($cf.$ \citealt{BH97}):

\begin{equation}\label{lick_mg_eq}
\textrm{Mg}_{2_{\sf{TTF}}} = -2.5 \,\, \textrm{log}_{10}  \left (
\frac{{\cal F}_{\sf{Mg}_{2}}}{{\cal F}_{c}}   \right )\, \textrm{[mag]} 
\end{equation}

\begin{equation}\label{lick_fe_eq}
\textrm{Fe}5270_{_{\sf{TTF}}} = \Delta \lambda \, \left ( 1 - 
\frac{{\cal F}_{\sf{Fe}5270}}{{\cal F}_{c}}   \right )\, [\AA] 
\end{equation}

\noindent
where ${\cal F}_{\sf{Mg}_{2}}$ and ${\cal F}_{\sf{Fe}5270}$ denotes
the sky-subtracted, filter profile-corrected flux at the {\mgtwo} and
Fe5270 line wavelengths, respectively, and $\Delta \lambda$ is the
FWHM of the Lorentzian bandpass.  The continuum flux, ${\cal F}_{c}$,
is given by:

\begin{equation}
{\cal F}_{c} = a \, {\cal F}_{RC} + b \, {\cal F}_{BC},
\end{equation}

\noindent
where $a + b = 1$, and ${\cal F}_{RC}$ and ${\cal F}_{BC}$ are the
sky-subtracted, filter profile-corrected flux at the red and blue
continua, respectively. We varied the weighting given to the red and
blue continua and found that the strongest correlation between our TTF
indices and the Lick values was obtained by using the red continuum
only (i.e. $a = 1$ and $b = 0$). Table~\ref{lickfit_table} presents
the gradients, y-intercepts, associated errors and chi-squared values
for the fit between TTF and Lick Fe5270 indices obtained using only
the red continuum (column 3), only the blue continuum (column 4), and
with equal weighting (column 5). Table~\ref{lickfit_table} only shows
the $a = 1$ case for {\mgtwo} (column 2) because our spectral scans
only extended down to $\sim$4870\,$\AA$ for about half of the stars;
however this case gave the best fit.  

The deterioration of the fit for the Fe5270 feature upon inclusion of
the blue continuum flux is largely due to our non-standard technique
for obtaining line core and continuum flux levels, which is more
sensitive to the gradient in the spectrum than the Lick/IDS
technique. The Lick/IDS Fe5270 blue continuum bandpass covers a narrow
wavelength range, just blueward of the absorption feature, that best
represents the true continuum level (see the thick horizontal lines in
Fig.~\ref{stellarscan_fig}).  By comparison, the Lick/IDS line and red
continuum bandpasses are 2.1$-$2.7 times wider than that for the blue
continuum. As evident from Fig.~\ref{stellarscan_fig}, the broad wings
of our Lorentzian transmission profile makes the blue continuum flux
measured by the TTF sensitive to the shape and gradient of the
spectrum around the 5220$-$5260 {\AA} range. Better results would be
expected if the TTF passband could be narrowed when switching from
line to blue continuum wavelengths. However, it was not possible to
alter the filter's plate spacings during charge-shuffle operations.

In the case of {\mgtwo}, observing the blue continuum introduces
further complications.  Firstly, measuring Mg$_{2,BC}\lambda4926$
involves changing from the f126 to the B4 blocking filter. Since
filters could not be switched while in charge-shuffle mode, one would
lose some of the advantage of this differential observing technique to
average over temporal variations in air mass and photometric
conditions during exposures. Secondly, the blue continuum range of
$4895.125 - 4957.625$\,\AA, as defined by Worthey et~al. (1994), is
contaminated in many active galaxies by [O\,{\sc iii}]~$\lambda4959$
emission. Our sample of eight galaxies contains three Seyferts, for
which this [O\,{\sc iii}] emission is likely to contribute to the flux
in the Mg$_{2,BC}$ images. For these reasons, and considering the fit
parameters shown in Table~\ref{lickfit_table}, the value ${\cal
F}_{c}$ in equations
\ref{lick_mg_eq} and
\ref{lick_fe_eq}, was determined exclusively from red continuum images
for both {\mgtwo} and Fe5270.

\begin{figure}
\centering
\includegraphics[width=8cm]{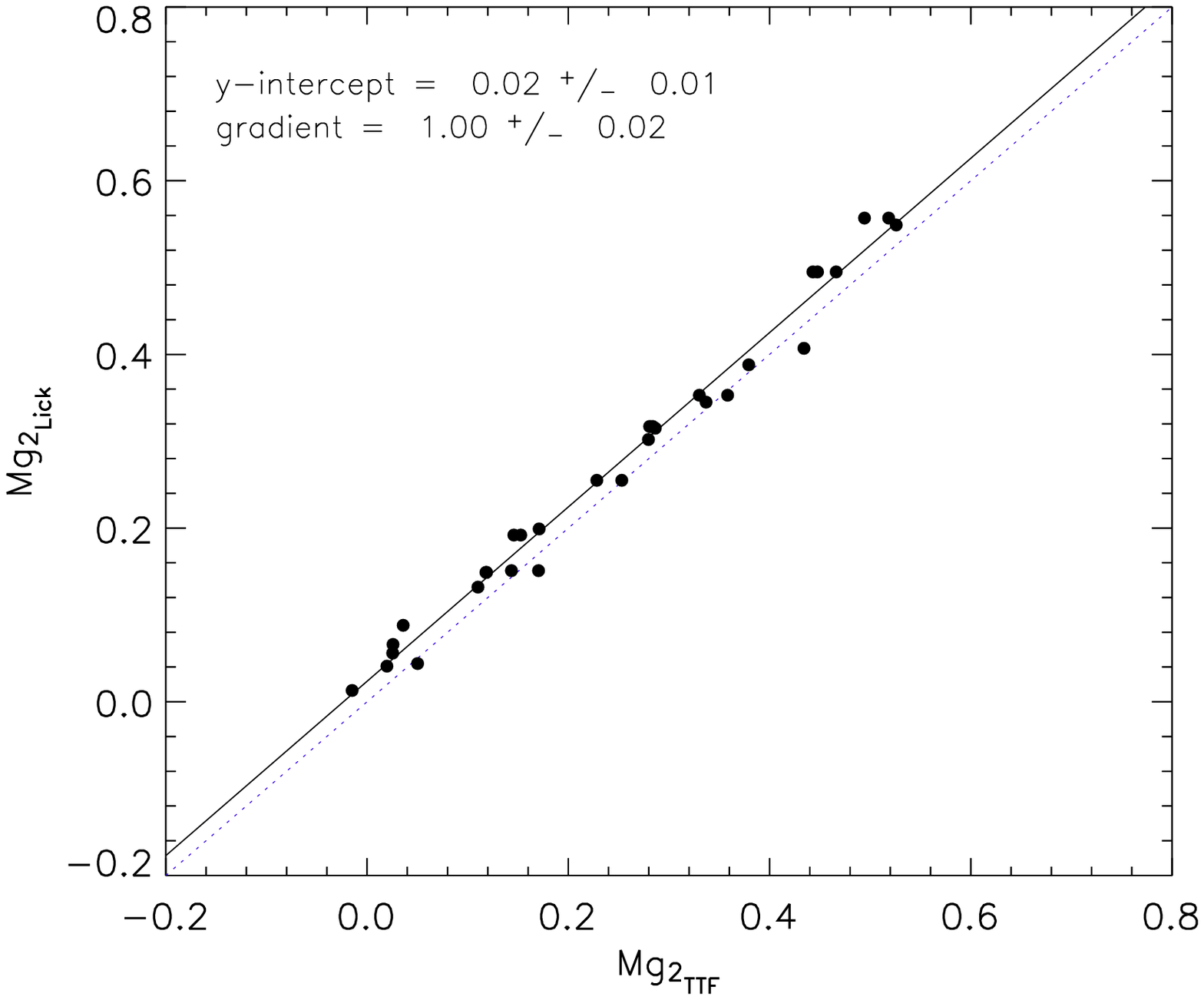}
\includegraphics[width=8cm]{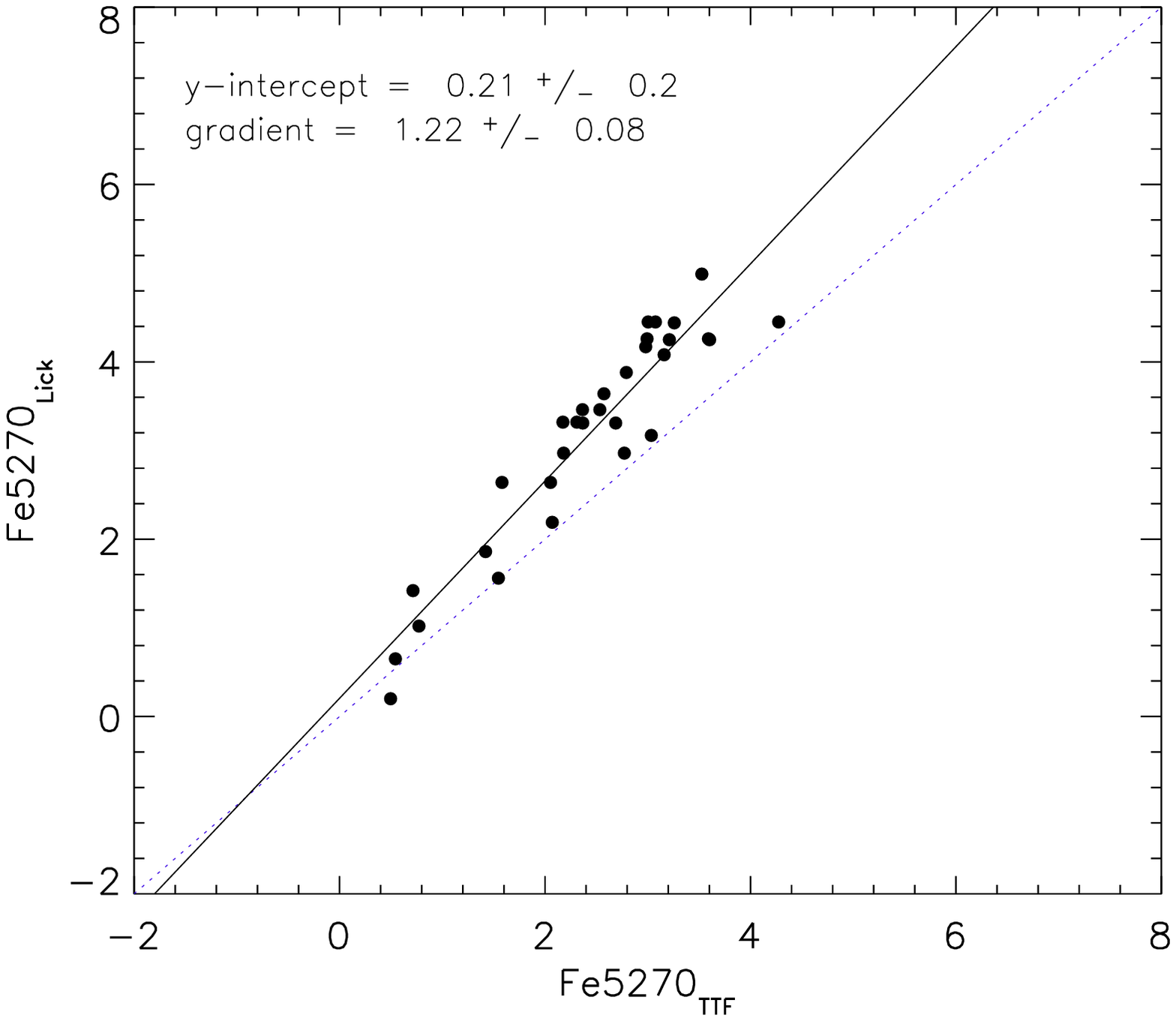}
      \caption{\emph{Top panel:} Relationship between measured TTF
      {\mgtwo} index and the published Lick {\mgtwo} value, where our
      TTF index was derived using the line and red continuum
      measurements only, as described in the text. Dotted line shows
      the one-to-one relationship. \emph{Bottom panel:} Same as top
      panel, but for the Fe5270 index.}
      \label{lickVSour3pt_paper_fig}
\end{figure}

Figure~\ref{lickVSour3pt_paper_fig} compares the {\mgtwo} and Fe5270
indices obtained for our sample of 31 stars using the absorption-line
imaging technique described in this study against the published Lick
values.  The following equations allow the transformation of
$\textrm{Mg}_{2_{\sf{TTF}}}$ and $\textrm{Fe}5270_{_{\sf{TTF}}}$ into
the Lick system, and were used for subsequent galaxy data analysis:

\begin{equation}\label{ttftolick_mg_eq}
\textrm{Mg}_{2_{\sf{Lick}}} = 1.00(\pm0.02) \,\, \textrm{Mg}_{2_{\sf{TTF}}} + 0.02(\pm0.01) \,\,\textrm{[mag]} 
\end{equation}

\begin{equation}\label{ttftolick_fe_eq}
\textrm{Fe}5270_{_{\sf{Lick}}} = 1.22(\pm 0.08) \,\,\textrm{Fe}5270_{_{\sf{TTF}}} + 0.2(\pm0.2)\, \,[\AA] 
\end{equation}

\begin{table}
\caption{Stellar Lick index fit parameters $^a$ \label{lickfit_table}}
\centering
\begin{tabular}{r|p{0.6cm} c p{0.9cm} p{0.9cm} l}
\hline
          & {\mgtwo}   &   &   & Fe5270 & \vspace{0.1cm}  \\
   \cline{2-2}   \cline{4-6}  \vspace{-0.2cm}   \\ 
${\cal F}$$_c$ $=$ &${\cal F}$$_{RC}$   & & ${\cal F}$$_{RC}$
        & ${\cal F}$$_{BC}$  & (${\cal F}$$_{RC}$$+$${\cal F}$$_{BC}$)/2 \\
\hline
gradient         & 1.00     & & 1.22   & 1.16 & 1.44    \\
$\Delta$gradient & 0.02     & & 0.08   & 0.24 & 0.13    \\
y-intercept      & 0.02     & & 0.2    & 1.5  & 0.4     \\
$\Delta$y-intercept & 0.01  & & 0.2    & 0.4  & 0.3     \\
chi-squared      & 0.02     & & 5.4    & 26   & 9       \\
\hline
\end{tabular}
\flushleft
\scriptsize{$^a$ for the equations $\textrm{Mg}_{2_{\sf{Lick}}}= \textrm{gradient} \times \textrm{Mg}_{2_{\sf{TTF}}}+ \textrm{y-intercept}$ and  $\textrm{Fe}5270_{_{\sf{Lick}}}= \textrm{gradient} \times \textrm{Fe}5270_{_{\sf{TTF}}}+ \textrm{y-intercept}$. }
\end{table}


\section{Galaxy Observations}\label{galobs}

Having established the credentials of our technique using Lick
standard stars, we observed a sample of eight spiral galaxies,
selected in order to satisfy the following needs: 1) to span a range
of morphologies, from early to late Hubble types; 2) to be roughly
face-on; and 3) to fill enough of our CCD such that the galactic disk
covers a large number of pixels but leaves sufficient area from which
to estimate the background sky level. The last criterion meant that
all our galaxies have diameters between about 2 and 4 arcmin. Choosing
galaxies with low inclination angles ensures that the stellar
rotational velocity width ``fits'' within the bandpass used in this
study. Some properties of the galaxies are summarised in
Table~\ref{gal_table}.

Galaxy observations consisted of 10~minutes of integration at each
wavelength, with the TTF passband being switched between the
redshifted line core and red continuum wavelengths each minute, while
synchronously shuffling the charge back and forth across the CCD
between an ``exposure'' and a ``storage'' location. Thus, temporal
variations in atmospheric and seeing conditions were smoothed out over
both spectral passbands (see Maloney \& Bland-Hawthorn 2001 for a
detailed description of this charge-shuffle technique). Multiple
charge-shuffled exposures were taken of each galaxy, with the total
exposure times for {\mgtwo} and Fe5270 given in
Table~\ref{obslog_table}.

\begin{table*}
\caption{Galaxy Properties \label{gal_table}}
\centering
\begin{tabular}{l|cccccccccc}
\hline
Galaxy & Hubble  & Recession &Distance & Inclination & m$_{B}$ & Scale & mean surf
& D$_{25}$  \\
      & Type$^{a}$  & Vel. (km s$^{-1}$)$^{a}$  & (Mpc)$^{b}$ &
  (deg)$^{c}$  &(mag)$^{a}$&(pc arcsec$^{-1}$) & brightness $^{c}$ &(arcmin)$^{c}$  \\
\hline
NGC 5968 & SAB(r)bc      & 5448 &  77     & 22  & 13.1  & 352 & 22.7 & 2.2\\
NGC 6221  & SB(s)bc pec  &  1482 &  20     & 62  & 10.7  & 96 &21.5   & 4.5   \\
NGC 6753 & (R')SA(r)b    &  3124 &  42      & 31  & 12.0  & 202 & 20.8 &  2.6  \\
NGC 6814  & SAB(rs)bc 	 &  1563 &  21    & 65  & 12.1  & 101 & 22.0 &  3.3  \\
NGC 6935 & (R)SA(r)a     &  4587 &  61     & 31  & 12.8  & 296 & 21.7 &  2.2  \\
NGC 7213  & SA(s)0 LINER &  1792 &  24     & 29  & 11.0  & 116& 20.7  &  3.0   \\
NGC 7412  & SAB(s)c	 &  1717 &  23     & 41  & 11.9  & 111 & 22.1 & 4.1  \\
NGC 7637 & SA(r)bc       &  3680 &  49      & 28  & 13.2  & 238 & 22.4 &  2.2  \\
\hline
\end{tabular}
\flushleft
\vspace{-0.4cm}
\begin{list}{}
\item {\scriptsize $^a$ de Vaucouleurs et al. 1991}
\vspace{-0.1cm}
\item {\scriptsize $^b$ Based on recession velocity and assuming a
Hubble constant of 75 km s$^{-1}$ Mpc$^{-1}$ }
\vspace{-0.1cm}
\item {\scriptsize $^c$ LEDA}
\end{list}

\end{table*}

\begin{table*}
\caption{Galaxy observation log \label{obslog_table}}
\centering
\begin{tabular}{l|rrcc}
\hline
Galaxy & Lick Feature & Date & Total Exp [min] & Notes  \\
\hline
NGC 5968  & Mg$_2$ line + R.C. & 31 July 2003  & 120 & {\scriptsize poor seeing and some clouds}   \\
	  & Fe5270 line + R.C. & 28 July 2003  & 60    \\
NGC 6221  & Mg$_2$ line + R.C. & 7 Aug 2002  & 120  & {\scriptsize poor seeing}    \\
          & Fe5270 line + R.C. & 6 Aug 2002  & 60    \\
NGC 6753  & Mg$_2$ line + R.C. & 29 July 2003  & 120    \\
	  & Fe5270 line + R.C. & 28 July 2003  & 60    \\
NGC 6814  & Mg$_2$ line + R.C. & 8 Aug 2002  & 40    \\
          & Fe5270 line + R.C. & 8 Aug 2002  & 60    \\
NGC 6935  & Mg$_2$ line + R.C. & 31 July 2003  & 120 &  {\scriptsize poor seeing}  \\
	  & Fe5270 line + R.C. & 28 July 2003  & 60    \\
NGC 7213  & Mg$_2$ line + R.C. & 7 Aug 2002  & 120  & {\scriptsize poor seeing and some clouds}   \\
          & Fe5270 line + R.C. & 6 Aug 2002  & 60    \\
NGC 7412  & Mg$_2$ line + R.C. & 7 Aug 2002  & 80 & {\scriptsize  some clouds}   \\
          & Fe5270 line + R.C. & 8 Aug 2002  & 60    \\
NGC 7637  & Mg$_2$ line + R.C. & 31 July 2003  & 100 & {\scriptsize poor seeing}   \\
	  & Fe5270 line + R.C. & 28 July 2003  & 60    \\
\hline
\end{tabular}
\end{table*}


\section{Galaxy data reduction and analysis}\label{datareductionanalysis}

In this section, we describe each of the steps involved in reducing
and analysing the galaxy data in order to evaluate the {\mgtwo} and
Fe5270 radial profiles.

\subsection{Data Reduction}\label{datareduction}
\subsubsection{Reduction of raw galaxy images}\label{galreduction}

Due to the charge-shuffle technique, the line and continuum images are
located in different regions on a single CCD image. After bias
subtraction, each image was bisected into the line and continuum
component. These images were trimmed for compatibility with the
flat-field images, which were not charge-shuffled.  Repeat galaxy
images were then averaged to produce four images per galaxy,
corresponding to the {\mgtwo} and the Fe5270 line and red continuum
wavelengths.

\subsubsection{Flat-field correction}\label{ff}

Correctly flat-fielding the galaxy images is crucial for eliminating
spatial variations in the pixel to pixel response due to filter
transmission and with wavelength/etalon spacing. Any residual
large-scale structure in the images will compromise subsequent surface
photometry.  After each night's observing, a series of dome flats was
taken, with the etalon set to the same spacings/wavelengths at which
the galaxies were observed. Typically, three exposures were taken per
wavelength, and combined into a single image, that was then normalised
and divided into the appropriate galaxy image, as part of the {\sc
iraf} flat-fielding procedures.

\subsubsection{Ellipse procedure}\label{ellipse}

The ELLIPSE package within {\sc iraf/stsdas} was used to analyse
separately, but identically, the radial surface brightness profiles of
the line and continuum images. There were four images per galaxy,
taken at the {\mgtwo} line and red continuum and Fe5270 line and red
continuum wavelengths. Position angle and ellipticity were fitted
interactively to one of the images, with the same settings being
applied to the remaining three images and fixed at all radii. Linear
radial steps were used and the average flux within each ellipse was
calculated with the ELLIPSE package using the median area integration
mode.  Discrepant data points were eliminated within ELLIPSE by
imposing 2- and 3-sigma upper and lower clipping criteria,
respectively, and by iterating the clipping algorithm four
times. H\,{\sc ii}~regions were not masked, but they are not expected
to significantly affect the measured surface brightness profile
because they generally occupy a small fraction of pixels per ellipse
and therefore should have a negligible influence on the median
flux. Moreover, the clipping algorithm removes many pixels associated
with bright H\,{\sc ii}~regions.

The ELLIPSE procedure yields radial surface brightness profiles for
{\mgtwo} and Fe5270 in the line and continuum for each galaxy. Each
line and continuum pair of surface brightness profiles then needed to
be sky-subtracted, filter-profile corrected, converted to instrumental
indices via equations~\ref{lick_mg_eq} and \ref{lick_fe_eq}, and finally
transformed into the Lick system using equations~\ref{ttftolick_mg_eq}
and \ref{ttftolick_fe_eq}. These steps are described in the following
section.

\subsection{Data Analysis}\label{analysis}

\subsubsection{Sky subtraction }\label{skysub}

The sky level was estimated from the mean flux in an elliptical
annulus whose inner radius and width was selected to ensure that the
radial variation in surface brightness was less than the
pixel-to-pixel dispersion in each ellipse. In this way, the sky was
taken from a ``flat'' part of the radial surface brightness profile
and, in most cases, the sky annulus lay well outside the isophotal
radius. For each galaxy, the same sky annulus was used for all four
{\mgtwo} and Fe5270 line and continuum images. The average brightness
in the sky annulus was then subtracted from the surface brightness
profile.

\subsubsection{Filter profile correction}\label{filterprofile}

Two methods of correcting for the wavelength-dependent filter
transmission profile were tested. In the first instance, we divided
each surface brightness profile by the value of the filter profile
(shown in Fig.~\ref{stellarscan_fig}) at the corresponding
wavelength, i.e. the same technique as for the standard stars.
However, better results were obtained by instead applying a scaling
factor to the radial profiles such that the value of the line and
continuum sky levels were equal. This forces the TTF instrumental
{\mgtwo} and Fe5270 indices to have an average value of zero in the
sky annuli.

There are several reasons why the second approach was more accurate
than the first approach, despite the former's success when used on
standard stars. Firstly, inspection of Fig.~\ref{stellarscan_fig}
reveals that the Fe5270 line and red continuum fall on a fairly flat
part of the f126 filter profile, such that filter transmission is not
very sensitive to drifts in wavelength calibration (that can be as
large as 4\,$\AA$, as will be discussed in Section~\ref{errors}
below). Although the {\mgtwo} line and red continuum fall on steeper
parts of the profile, small shifts in wavelength calibration would
change the transmission of the line and continuum by similar
amounts. Thus, the \emph{relative} transmission of the line and
continuum features for both {\mgtwo} and Fe5270 is fairly robust to
wavelength calibration drifts \emph{for objects at or near
rest-wavelength}. The recessional velocities of our sample of galaxies
(given in Table~\ref{gal_table}) shift the spectral features toward
longer wavelengths by 30 -- 100\,$\AA$. This tends to place the galaxy
line and continua in regions of the spectrum where their
\emph{relative} transmission is far more sensitive to small wavelength
drifts. Secondly, unlike the stellar observations, the galaxy
exposures lasted at least an hour, during which time the TTF
alternated between line and continuum wavelengths at one minute
intervals. As a result, the amount of light reaching the CCD varied
smoothly with changing airmass but also varied unpredictably with
fluctuating photometric conditions. If one applies the filter profile
correction factor to the surface brightness profiles, it is still
necessary to scale the images to account for residual differences in
atmospheric conditions. Conversely, by using the relative line and red
continuum sky levels to correct for filter transmission, one naturally
corrects for variable observing conditions during exposures. However,
this approach assumes that the intensity of the background sky is the
same at both the line and red continuum wavelengths.

\subsubsection{Conversion to Lick system}\label{conversion}

Finally, equations~\ref{lick_mg_eq} and \ref{lick_fe_eq} were used to
derive instrumental indices and equations~\ref{ttftolick_mg_eq} and
\ref{ttftolick_fe_eq} allowed the transformation into the Lick system. 

\subsection{Errors}\label{errors}

Our results are subject to various sources of uncertainty, the most
important of which are: 1) sky subtraction, 2) filter transmission
profile correction, and 3) TTF to Lick conversion. Each of these
sources of error act on the calculated radial line-strength profiles
in different ways. The foremost factor influencing the line-strength
\emph{gradients} is the estimation and subtraction of the background
sky level, which introduces very large errors toward the outer
galactic disk, where the galaxy surface brightness is very low
relative to the sky. The sky subtraction uncertainty, $\sigma_{sky}$,
was evaluated from the flux dispersion in the annulus designated as
the sky region. The position of this sky annulus was somewhat
subjective and was chosen to minimise the rms dispersion.

The error, $\sigma_{filter}$, due to correcting for the
wavelength-dependent filter transmission profile does not influence
the shape or gradient of the radial line-strength profile, but instead
can uniformly shift the radial profile vertically.

Converting the instrumental indices into the Lick system introduces
the errors specified in equations \ref{ttftolick_mg_eq} and
\ref{ttftolick_fe_eq}. These errors affect both the profile shape and
the absolute values of the line-strengths. It should be emphasised
that the uncertainty in the profile shape due to sky subtraction
overwhelms that associated with TTF-to-Lick conversion.

Although we frequently re-calibrated our $Z(\lambda)$ relation between
galaxy exposures, the wavelength tuning was found to drift by up to
5\,$\AA$ over the course of a night. Such a drift introduces errors
into the measured indices in two ways: firstly, one loses sensitivity
to the absorption-line as the bandpass shifts away from the centroid
of the absorption feature; and secondly, the blocking filter
transmission profile is strongly wavelength-dependent - especially for
$\lambda$ $\lesssim 5230\,\AA$ and $\gtrsim 5300\,\AA$, as evident in
Fig.~\ref{stellarscan_fig}. The error introduced by this wavelength
calibration drift is, however, already folded into the TTF-to-Lick
conversion error. This is because the stellar observations used to
transform TTF instrumental indices into the Lick system were
themselves subject to calibration drift, and this contributes to the
scatter seen in Fig. \ref{lickVSour3pt_paper_fig}.

Seeing varied from less than 1 arcsec to about 4 arcsec in the worst
conditions, but because the flux was azimuthally averaged in $\sim$ 5
arcsec spaced radial bins, poor seeing is only likely to have affected
the calculated line-strengths in the innermost radii. Moreover, since
the radial gradients listed in Table~\ref{gal_results_tab} refer to
the disk only, excluding the central bulge, our calculations are
barely affected by changing seeing conditions.


\section{Results}\label{results}

Figures~\ref{5968_fig}$-$\ref{7637_fig} illustrate the main results of
this study. Images of each galaxy are presented along with plots of
the radial variation of surface brightness and the {\mgtwo} and Fe5270
indices. In each figure a 288 $\times$ 288 arcsec image of the galaxy
in the Fe5270 line is presented in panel $(a)$, overlaid with two
ellipses indicating the inner and outer bounds of the region from
which the background sky level was calculated. For each individual
galaxy, the ellipse geometry was fitted using the technique described
in Section~\ref{ellipse} and fixed for all radii.  Panel $(b)$ in each
figure shows the radial variation in surface brightness, with the
thick solid line denoting the location of the background sky
annulus. Panels $(c)$ and $(d)$ show the radial gradients of the
{\mgtwo} and Fe5270 Lick indices, respectively, calculated using the
method described in Section~\ref{datareductionanalysis}. The data in
panels $(c)$ and $(d)$ are tabulated in Appendix~A. Note that the
surface brightness profile plots extend to greater radii than the line
strength gradient plots. The total uncertainty due to sky subtraction,
filter transmission correction and transformation into the Lick system
is shown by thin error bars, while the thick bars display the error
due to sky subtraction alone. The dashed line indicates the sky
subtraction error-weighted line of best fit to the disk. Since the
radial gradients are most sensitive to uncertainty in sky subtraction,
they were calculated using only sky subtraction errors to weight the
data points.  As pointed out in Section~\ref{errors}, the errors from
filter transmission correction and transformation into the Lick system
act to uniformly shift the curves vertically, but do not change the
gradient. The disk gradient calculations excluded the inner bulge,
which was identified from the change in slope of the radial
log(surface brightness) profile, indicative of a bulge-to-disk
transition.

Table~\ref{gal_results_tab} summarises the {\mgtwo} and Fe5270 indices
derived for the inner bulge along with the gradient and vertical
intercept of the linear fit to the radial line-strength profiles in
the disk of the eight galaxies in our sample. The disk gradients
exclude the galactic bulge and are expressed with respect to the
apparent radius, R$_{25}$\,=\,D$_{25}/2$, where D$_{25}$ is listed in
Table~\ref{gal_table}.

One motivation for this kind of study is to test the dependence of
abundance gradients in disk galaxies on Hubble type. In an
investigation into line-strength gradients in S0 galaxies, Fisher,
Franx \& Illingworth (1996) found evidence for a strong relationship,
with the shallowest gradients being found in S0 galaxies and the
steepest in later-type spirals. We find no clear trend between Hubble
type and gradient in our small sample, but we note that the only S0
galaxy in our sample also has the only positive gradient.
Unfortunately, our sample is too small to draw statistically
significant conclusions.

As this study presents the first ever measurements of Lick indices
across the disks of these eight galaxies, a direct comparison of our
main results with the literature was not possible.  Restrictions
imposed by the availability of targets meeting the selection criteria
outlined in Section~\ref{galobs} prevented any overlap between our
sample and that of \citet{BH97} and \citet{MHB99}. \citet{IdFPC96}
measured Lick indices in the bulge of one of our galaxies, NGC~6935,
finding {\mgtwo} = 0.243\,$\pm$\,0.009\,mag and Fe5270 =
2.3\,$\pm$\,0.23\,{\AA}.  Reassuringly, our value of Fe5270 =
2.4\,$\pm$\,0.4\,{\AA} in the central galaxy is in very good agreement
with Idiart et~al. (1996), although our value of Mg$_{2}$ =
0.29\,$\pm$\,0.03\,mag is slightly higher.  We note that, unlike
Idiart et~al. (1996), our values have not been corrected for velocity
dispersion, nor reddening. However, \citet{BH97} demonstrated that
spectral broadening due to stellar velocity dispersion is small for
face-on spirals like those in our sample. From the results of
\citet{BH97}, we also expect that any reddening corrections are less
than uncertainties in the data and would not significantly affect the
index gradients. We emphasise that the chief sources of error
affecting the \emph{absolute} values of our derived Mg$_2$ and Fe5270
in the bulge come from correcting for the filter transmission profile
and transforming into the Lick system. For each galaxy, these
uncertainties translate into roughly the same offset in absolute value
for all radii. The following discussions concentrate on the radial
behaviour and \emph{gradients} of the line-strengths: properties that
are robust to most sources of uncertainty aside from sky subtraction.

Cid Fernandes, Storchi-Bergmann \& Schmitt (1998) measured radial
variations in stellar absorption features in the inner regions of
active galaxies, including the three Seyfert galaxies in our sample;
NGC~6221, NGC~6814 and NGC~7213. They did not use the Lick system to
measure spectral lines, however their radial profiles of the
equivalent width (EW) of the Mg~I+Mg~H feature allow for a qualitative
comparison with our Mg$_2$ measurements. Their wavelength window of
5156$-$5196$\,\AA$ for Mg~I+Mg~H matches the Lick definition for
{\mgtwo}, but Cid Fernandes et~al. (1998) use different
pseudo-continua. As will be discussed in more detail below, the shape
of their Mg~I+Mg~H profiles compare well with our Mg$_2$ profiles for
all three galaxies, although their measurements only extend to radii
less than a quarter those reached in this investigation.

\begin{table*}
\caption{Central indices and disk gradients \label{gal_results_tab}}
\centering
\begin{tabular}{l|cccclc}
\hline
Galaxy & Mg$_{2}$(r\,=\,0)  & Fe5270(r\,=\,0) & $\Delta$Mg$_{2}$\,/\,R$_{25}$ $^a$ & 
$\Delta$Fe5270\,/\,R$_{25}$ $^a$ & Mg$_{2}$ intercept $^b$ &  Fe5270 intercept $^b$  \\
\hline
NGC 5968 & 0.22 $\pm$ 0.03  & 3.2 $\pm$ 0.4 & $-$0.15 $\pm$ 0.05 & $-$0.8 $\pm$ 0.7 & 0.26 $\pm$ 0.02  & 3.0 $\pm$  0.3  \\
NGC 6221 & 0.17 $\pm$ 0.03  & 1.9 $\pm$ 0.3 & 0.0 $\pm$ 0.06     & $-$1.5 $\pm$ 1   & 0.17 $\pm$ 0.02  & 1.6 $\pm$  0.3  \\
NGC 6753 & 0.40 $\pm$ 0.03  & 2.6 $\pm$ 0.4 & $-$0.09 $\pm$ 0.02 & $-$1 $\pm$ 0.5   & 0.268 $\pm$ 0.005 & 2.8 $\pm$  0.1  \\
NGC 6814 & 0.23 $\pm$ 0.03  & 1.0 $\pm$ 0.3 & $-$0.1 $\pm$ 0.05  & $-$2 $\pm$ 1     & 0.27 $\pm$ 0.01  & 2.3 $\pm$  0.2  \\
NGC 6935 & 0.29 $\pm$ 0.03  & 2.4 $\pm$ 0.4 & $-$0.15 $\pm$ 0.03 & $-$1.7 $\pm$ 0.4 & 0.32 $\pm$ 0.01  & 2.8 $\pm$  0.1  \\
NGC 7213 & 0.23 $\pm$ 0.03  & 1.7 $\pm$ 0.3 & 0.15 $\pm$ 0.04    & 2  $\pm$ 1       & 0.15 $\pm$ 0.01  & 1.1 $\pm$  0.2  \\
NGC 7412 & 0.16 $\pm$ 0.03  & 2.0 $\pm$ 0.4 & $-$0.05 $\pm$ 0.1  & $-$1 $\pm$ 1.5   & 0.15 $\pm$ 0.02  & 2.0 $\pm$  0.2  \\
NGC 7637 & 0.24 $\pm$ 0.03  & 2.2 $\pm$ 0.3 & $-$0.15 $\pm$ 0.03 & $-$1.5 $\pm$ 0.5 & 0.253 $\pm$ 0.005 & 2.9 $\pm$  0.1  \\
\hline
\end{tabular}
\flushleft
\vspace{-0.4cm}
\begin{list}{}
\item {\scriptsize $^a$ Disk gradient (excluding inner bulge),
expressed with respect to the apparent radius, R$_{25}$=D$_{25}$/2, listed in
Table~\ref{gal_table}}.
\vspace{-0.1cm}
\item {\scriptsize $^b$ Vertical intercept of the linear fit to the disk
radial line-strength profiles - i.e. the extrapolation to zero radius
of the dashed lines seen in the panels $b)$ and $c)$ of
Figs.~\ref{5968_fig}$-$\ref{7637_fig}}.
\end{list}

\end{table*}

\section{Individual Galaxies}\label{individual_results}

In this section, we comment on the radial line-strength profiles for
each galaxy individually.

\subsection{NGC 5968 - SAB(r)bc}\label{ngc5968} 

\begin{figure*}
\centering \vspace{30pt}
  \hbox{\hspace{0.5in} (a) \hspace{3.4in} (c)} 
  \centerline{\hbox{ \hspace{0.8in} 
\includegraphics[width=6cm]{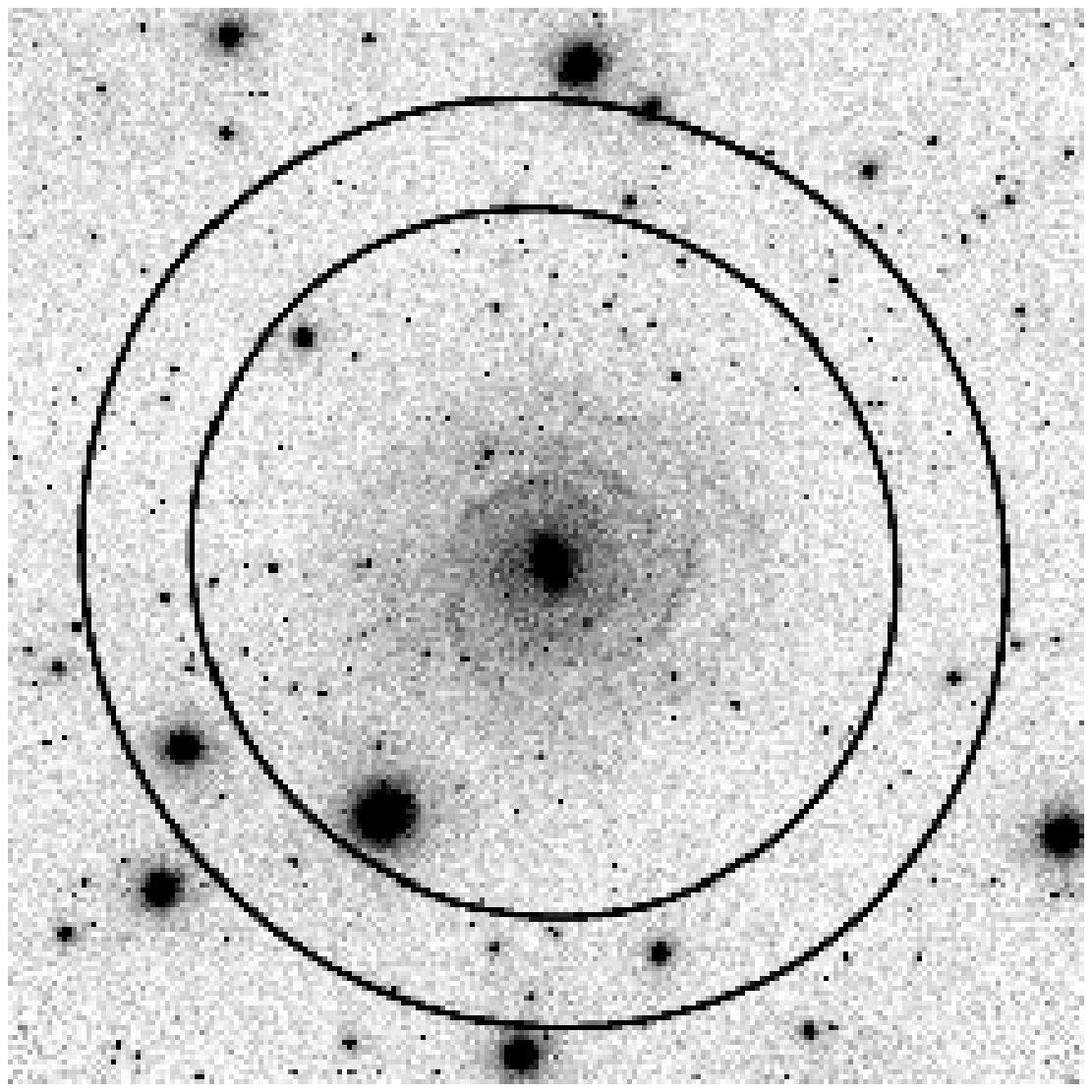}
    \hspace{0.3in}
\includegraphics[width=9cm]{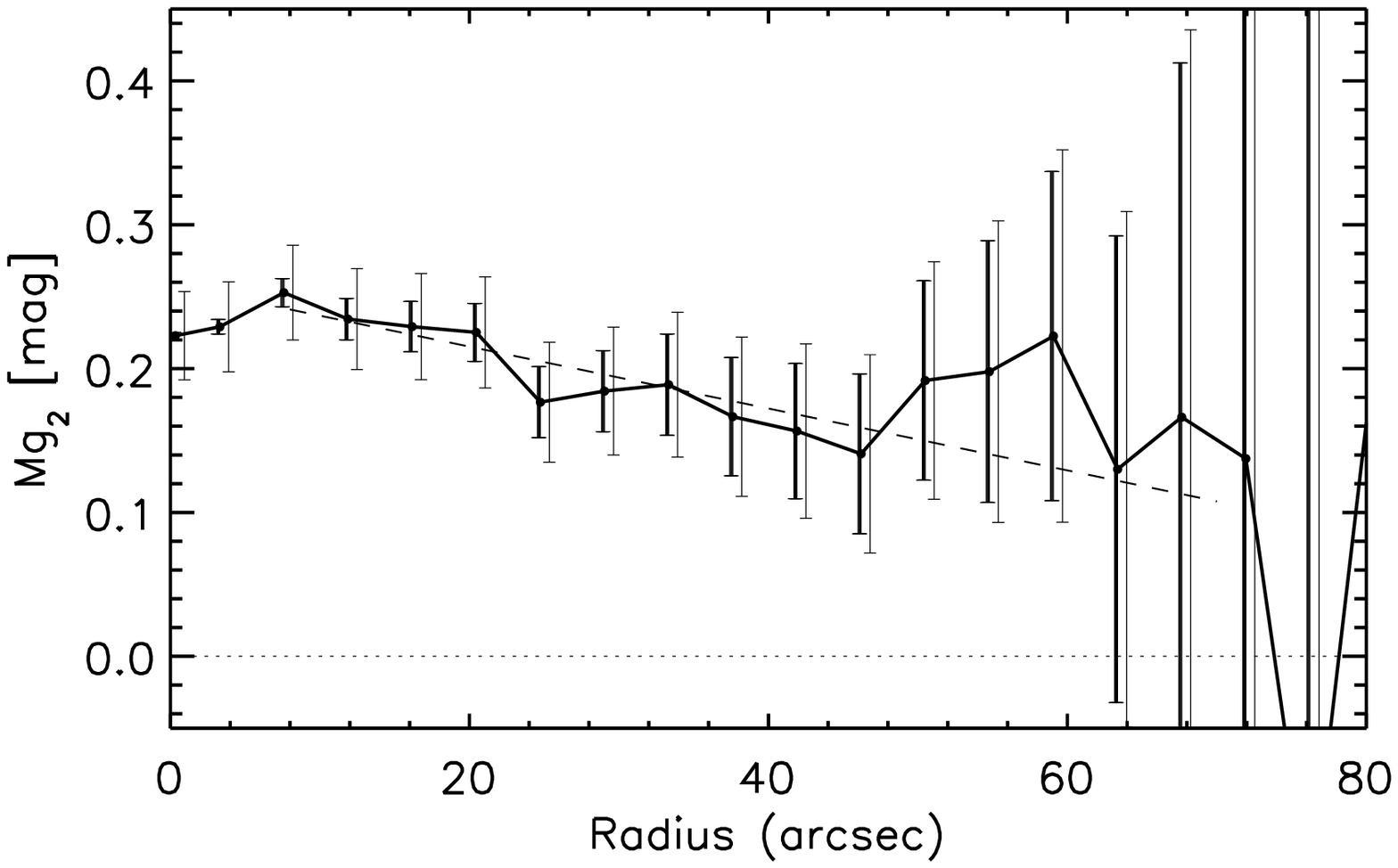}
    }
  }
  \vspace{30pt}
  \hbox{\hspace{0.5in} (b) \hspace{3.4in} (d)} 
  \centerline{\hbox{
\includegraphics[width=9cm]{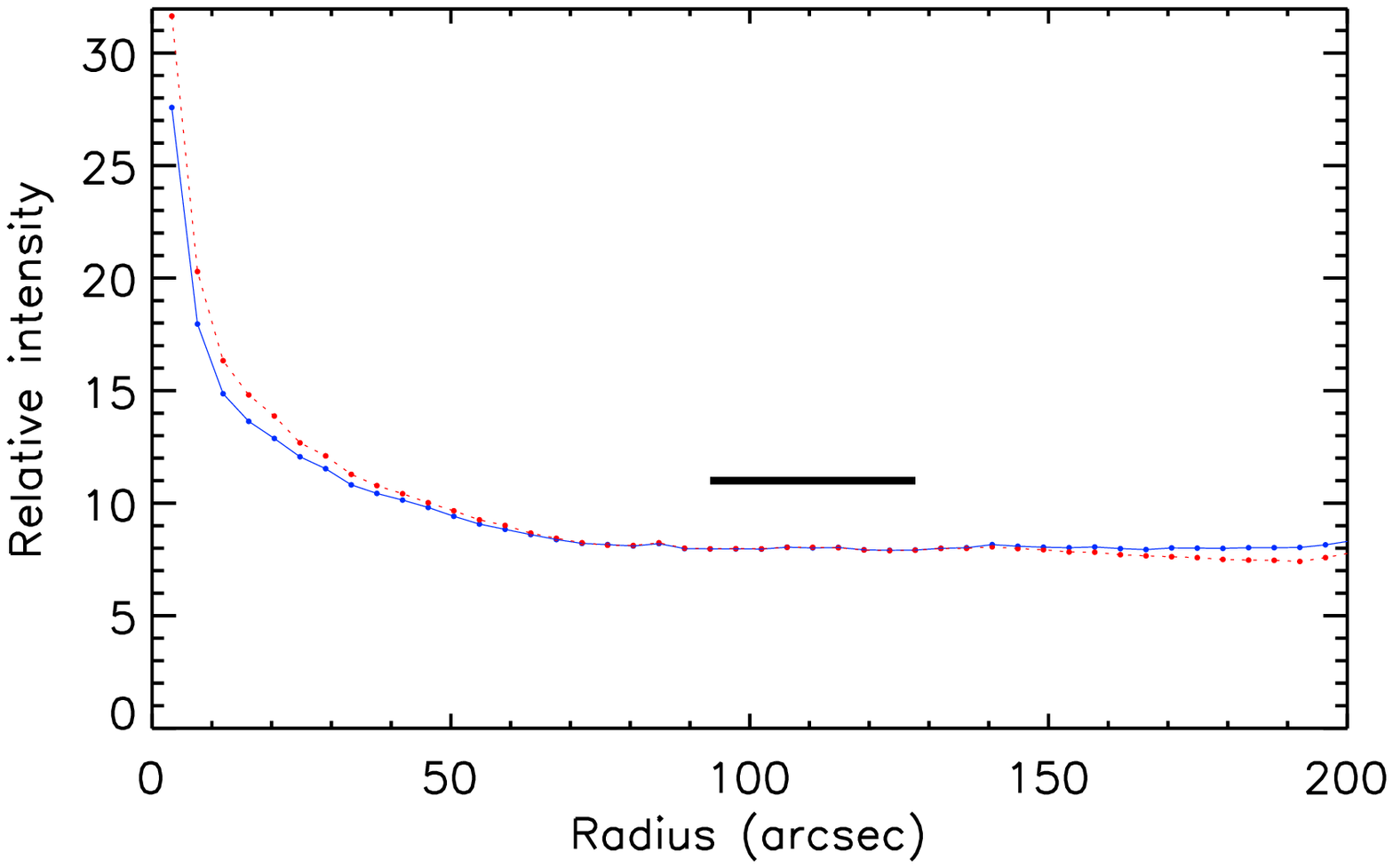}
\includegraphics[width=9cm]{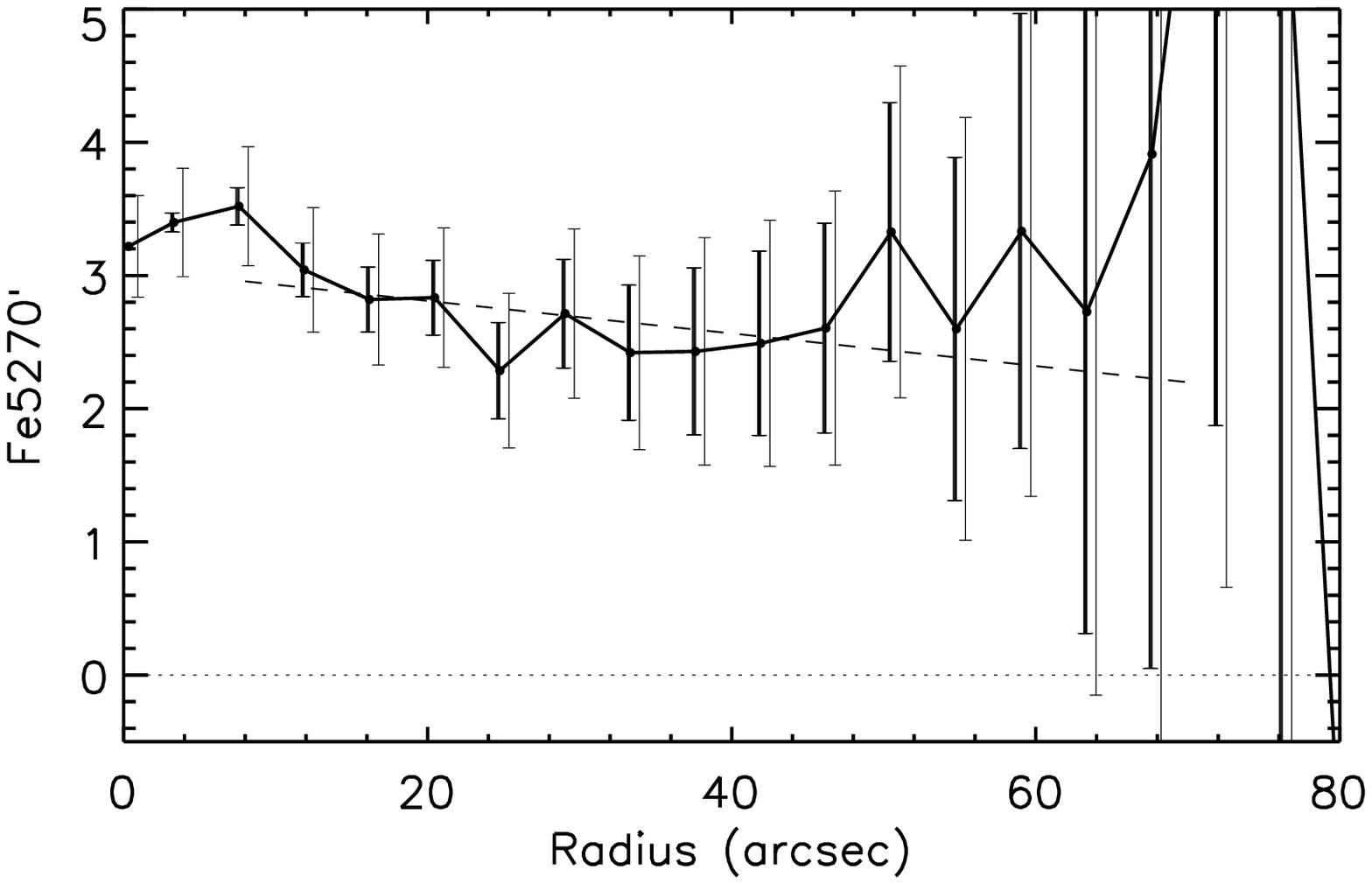}
    } } \vspace{9pt} \caption{NGC 5968: a) 288 $\times$ 288 arcsec
    image of the galaxy in the Fe5270 line. The two overlaid ellipses
    indicate the inner and outer bounds of the area from which the
    background sky level was estimated (see text for details); b)
    Surface brightness profile in the Mg$_{2}$ line (solid curve) and
    continuum (dotted curve). The thick straight line indicates the
    annulus from where the background sky level was estimated and the
    dots show the location of each ellipse; c) Mg$_2$ index,
    transformed into the Lick system using
    equation~\ref{ttftolick_mg_eq}. Two error bars are plotted for
    each data point. The thin bars show total error, including
    uncertainty associated with sky subtraction, filter transmission
    correction and transformation into the Lick system. The thick
    error bars show the contribution from sky subtraction only. The
    dashed line is the line of best fit to the data for the disk only
    (excluding the inner bulge). In deriving the line-strength
    gradient, only the error from sky subtraction was used to weight
    the data points, since this is the chief source of uncertainty
    affecting the measured gradient. Errors due to filter transmission
    correction and transformation into the Lick system act to
    uniformly shift the curve vertically, but do not change the
    gradient; d) Fe5270 index, transformed into the Lick system using
    equation~\ref{ttftolick_fe_eq}. Error bars and dashed line have
    the same meaning as in panel $c)$.}  \label{5968_fig}
\end{figure*}

\begin{figure*}
\centering \vspace{30pt}
  \hbox{\hspace{0.5in} (a) \hspace{3.4in} (c)} 
  \centerline{\hbox{ \hspace{0.8in} 
\includegraphics[width=6cm]{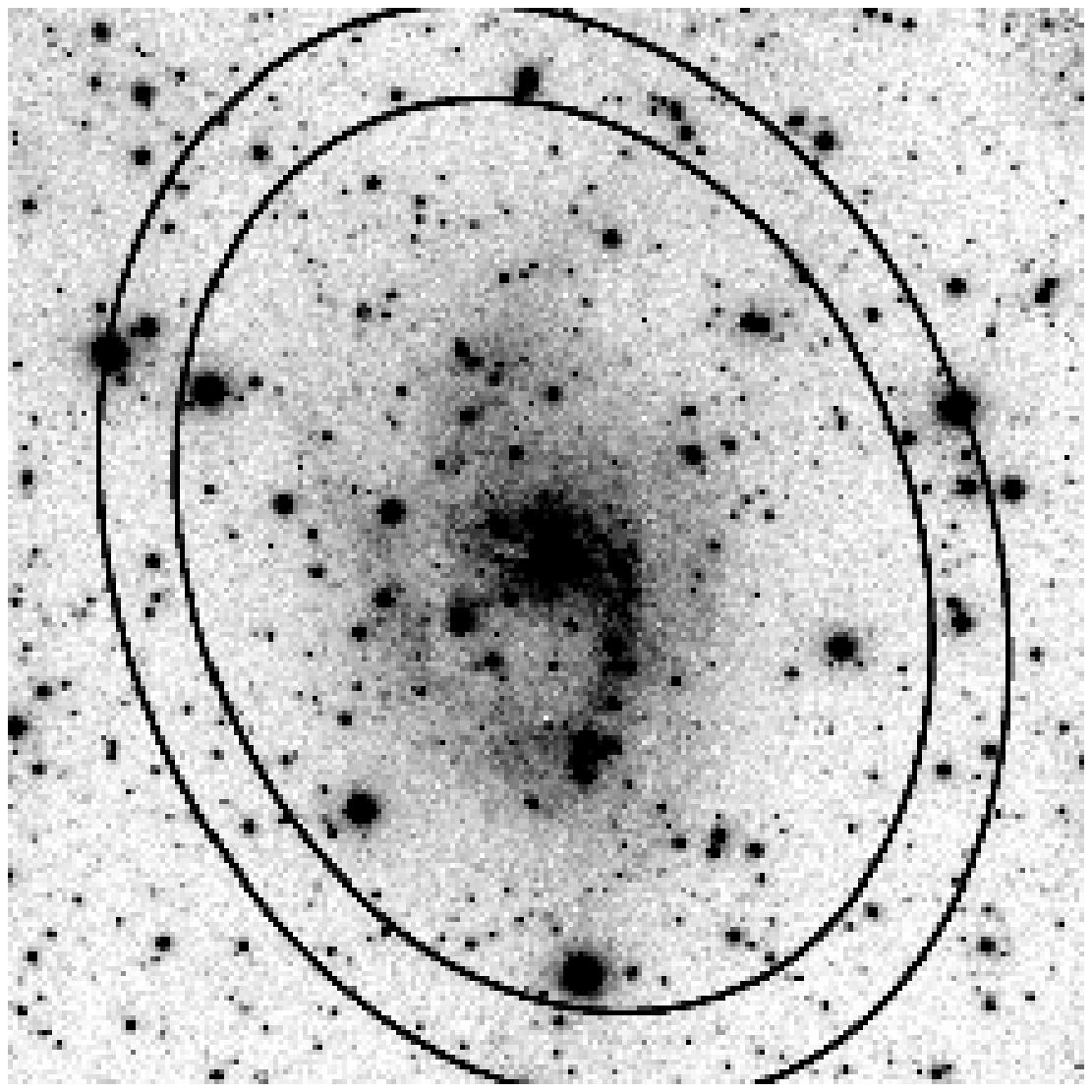}
    \hspace{0.3in}
\includegraphics[width=9cm]{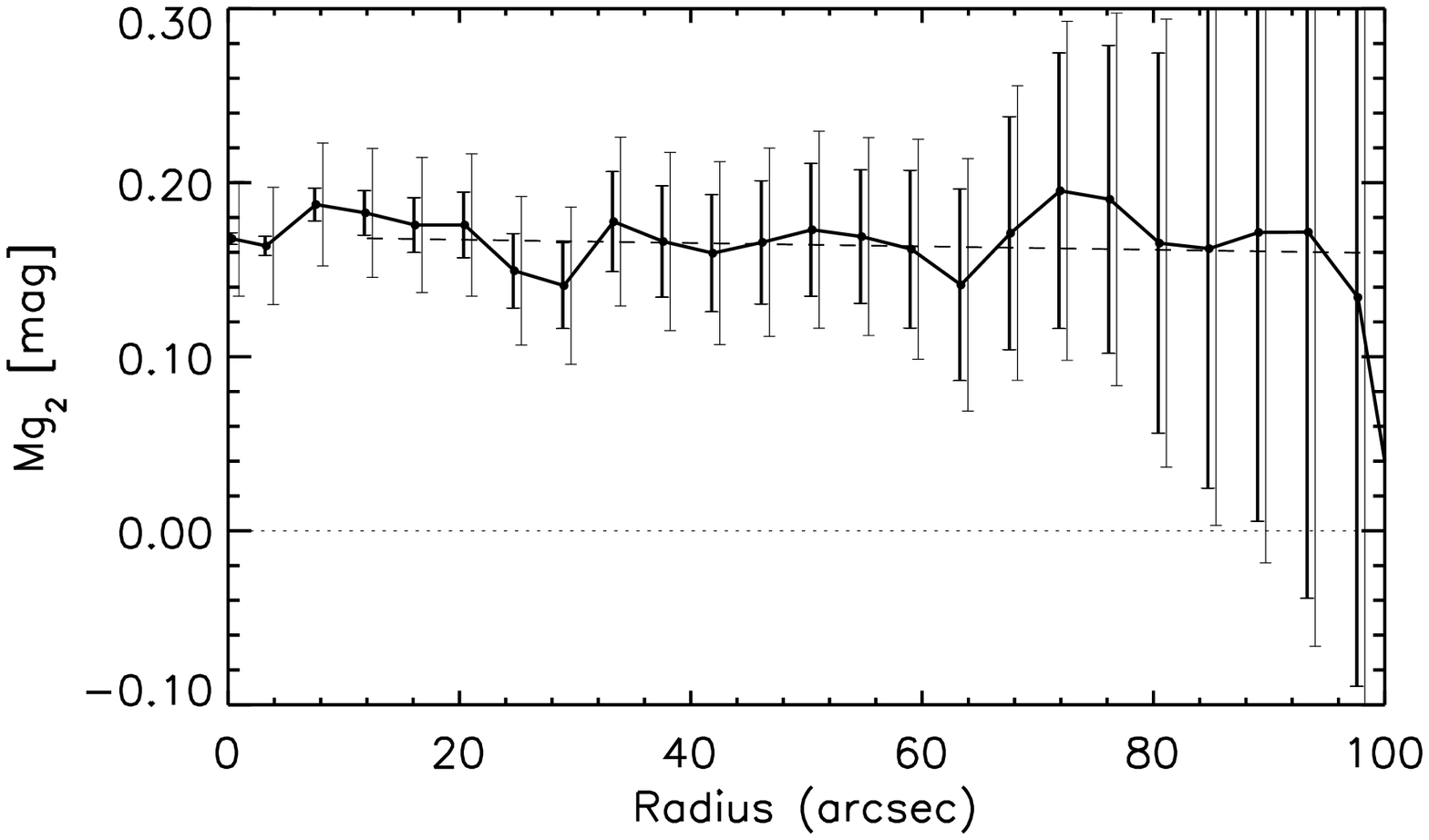}
    }
  }
  \vspace{30pt}
  \hbox{\hspace{0.5in} (b) \hspace{3.4in} (d)} 
  \centerline{\hbox{
\includegraphics[width=9cm]{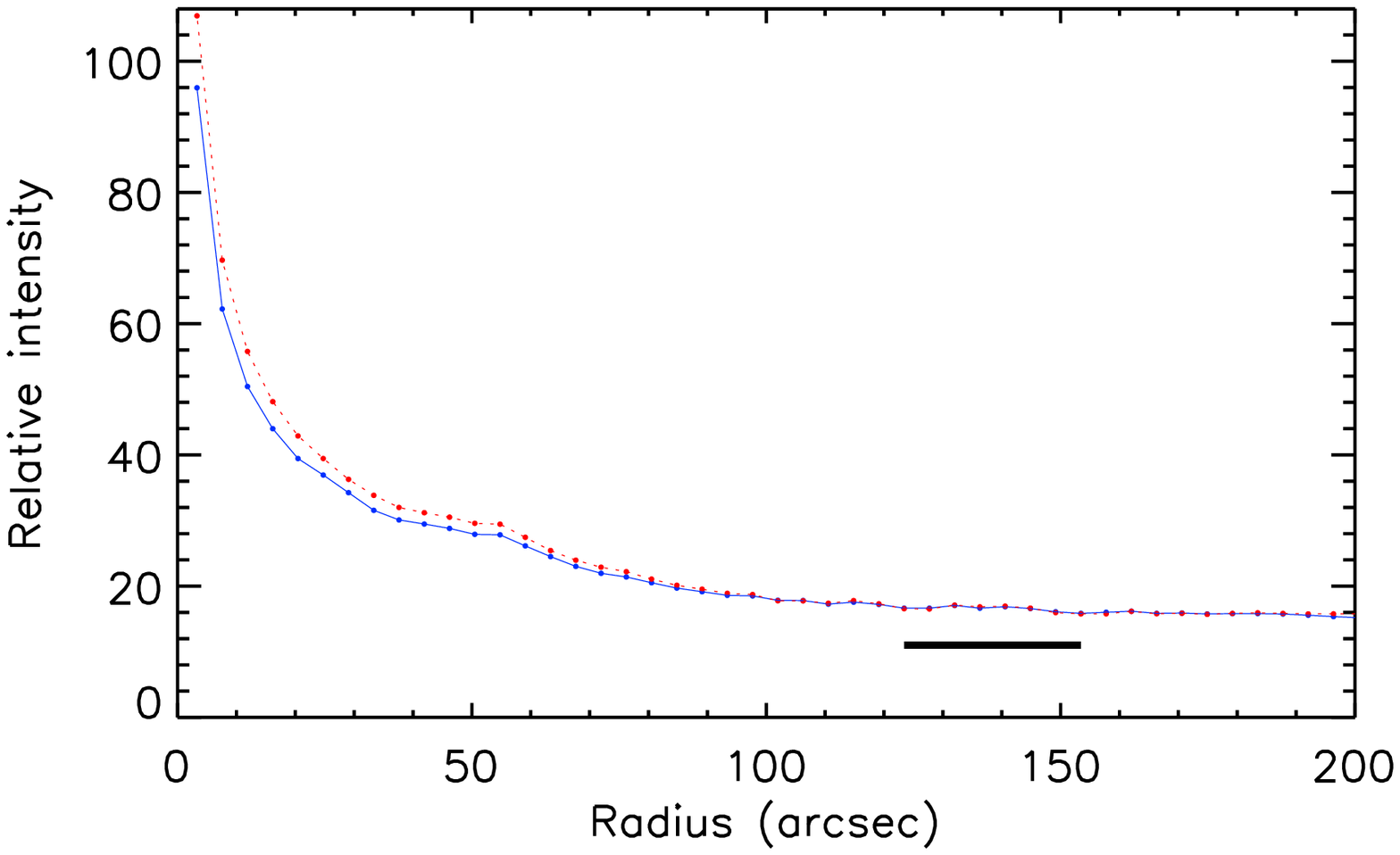}
\includegraphics[width=9cm]{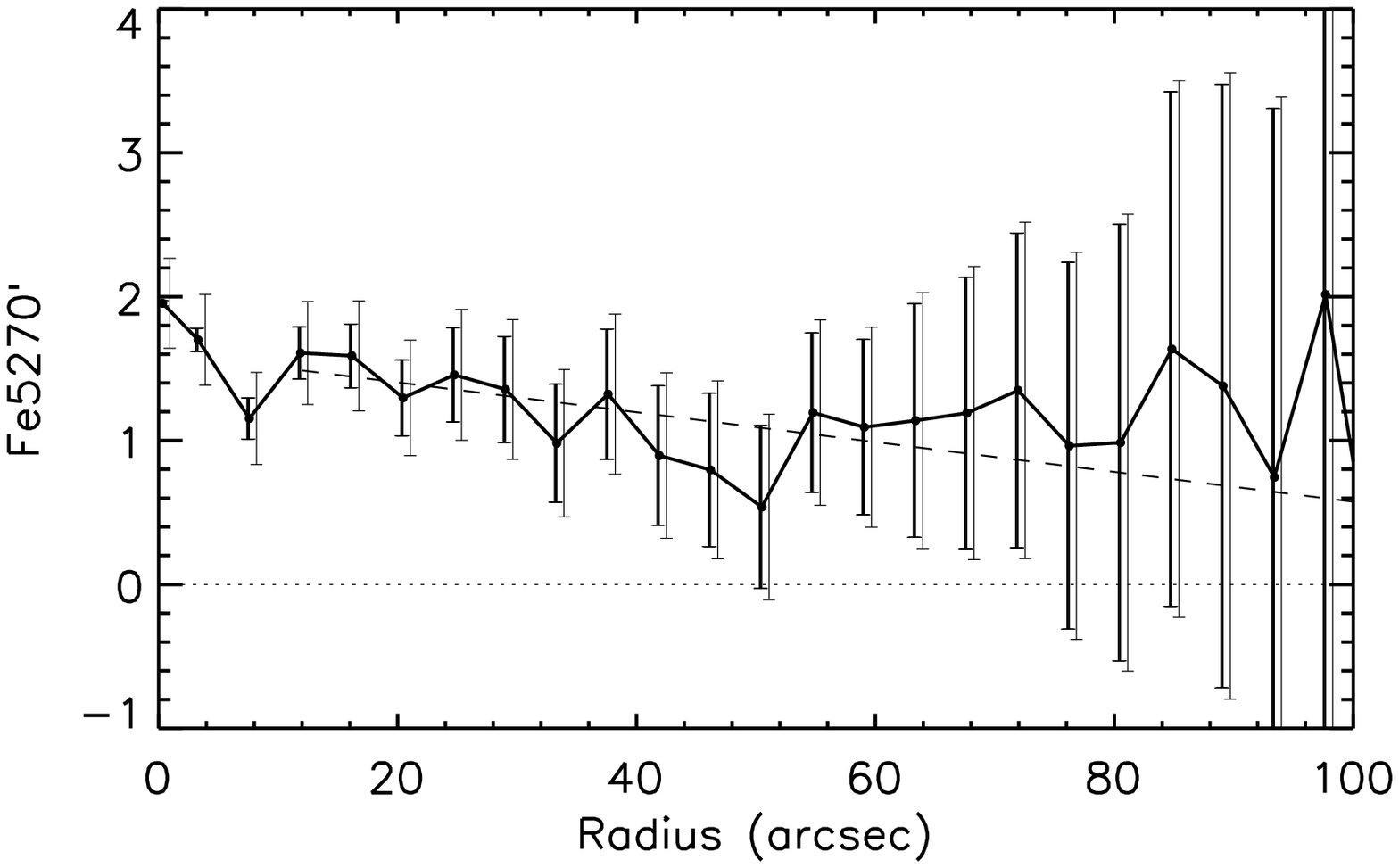}
    }
  }
  \vspace{9pt}
  \caption{Same as Fig.~\ref{5968_fig} but for NGC 6221}
  \label{6221_fig}
\end{figure*}

\begin{figure*}
\centering \vspace{30pt}
  \hbox{\hspace{0.5in} (a) \hspace{3.4in} (c)} 
  \centerline{\hbox{ \hspace{0.8in} 
\includegraphics[width=6cm]{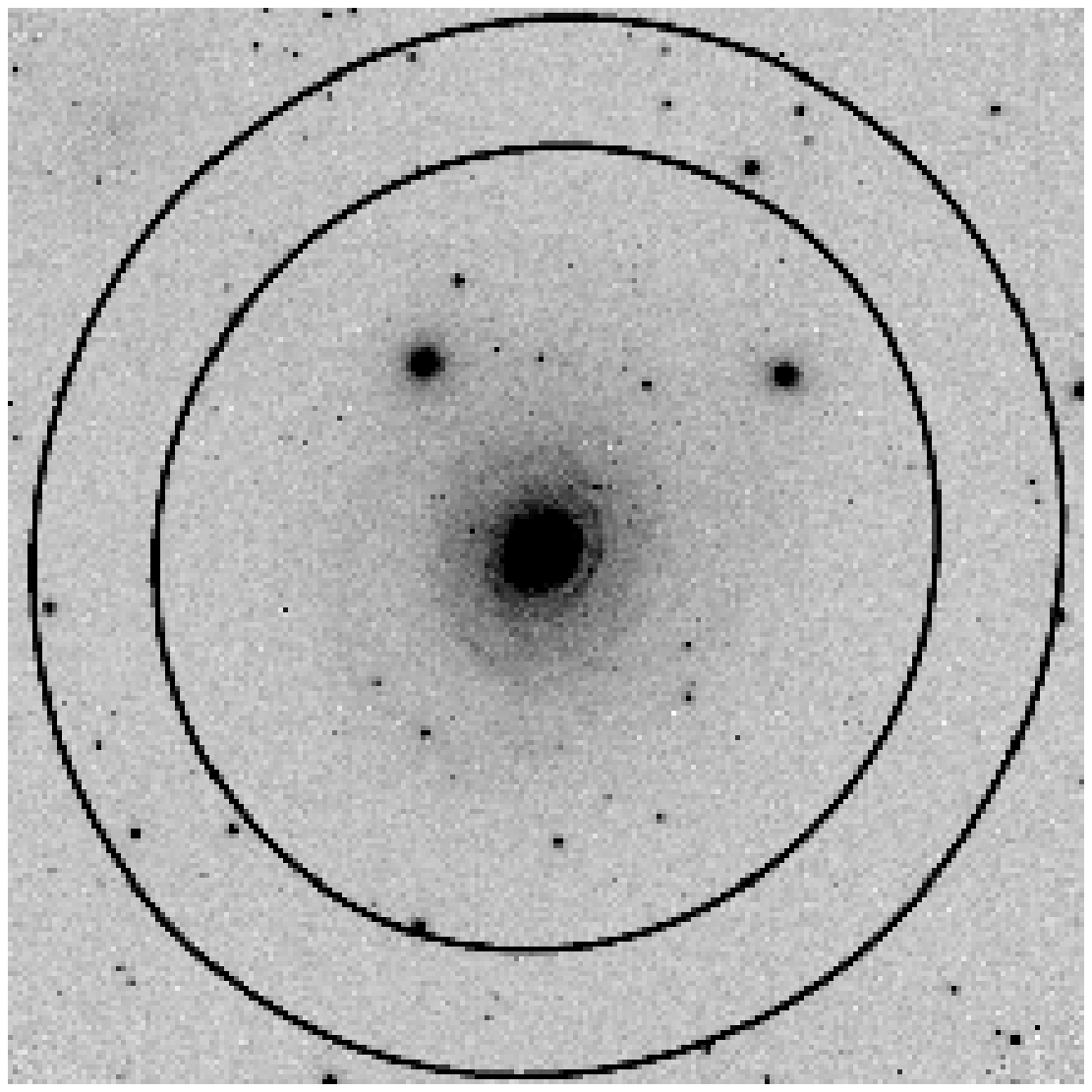}
    \hspace{0.3in}
\includegraphics[width=9cm]{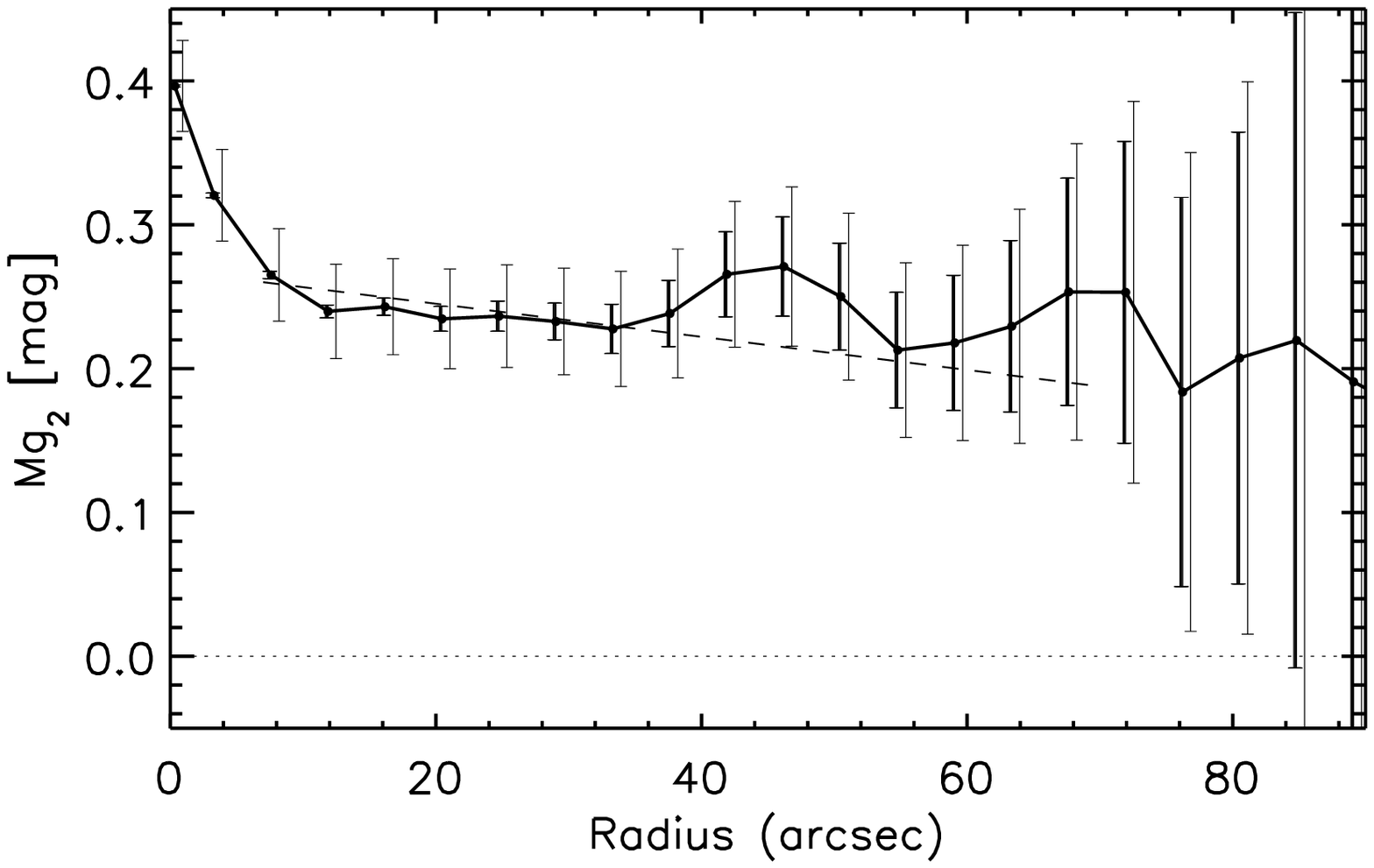}
    }
  }
  \vspace{30pt}
  \hbox{\hspace{0.5in} (b) \hspace{3.4in} (d)} 
  \centerline{\hbox{
\includegraphics[width=9cm]{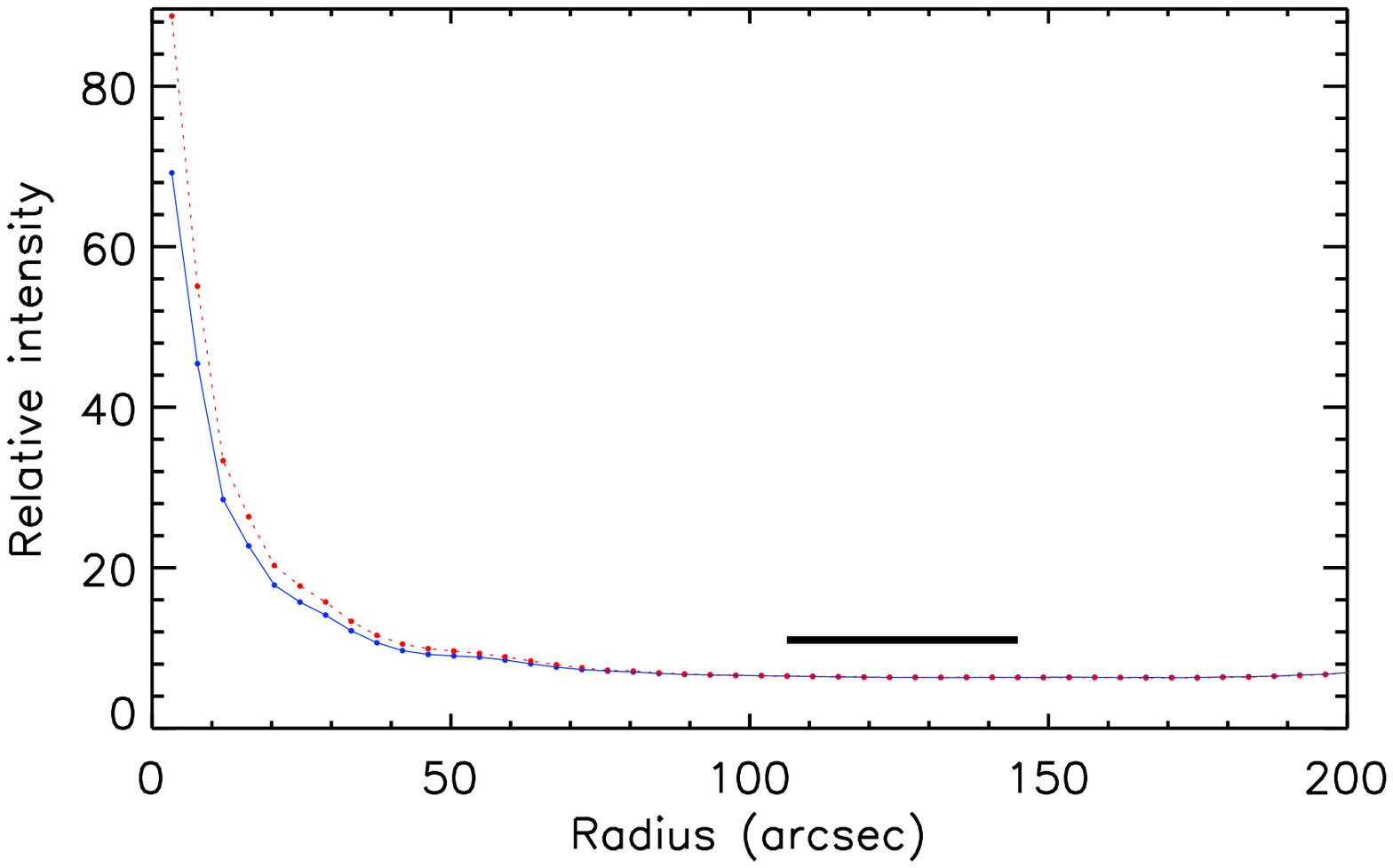}
\includegraphics[width=9cm]{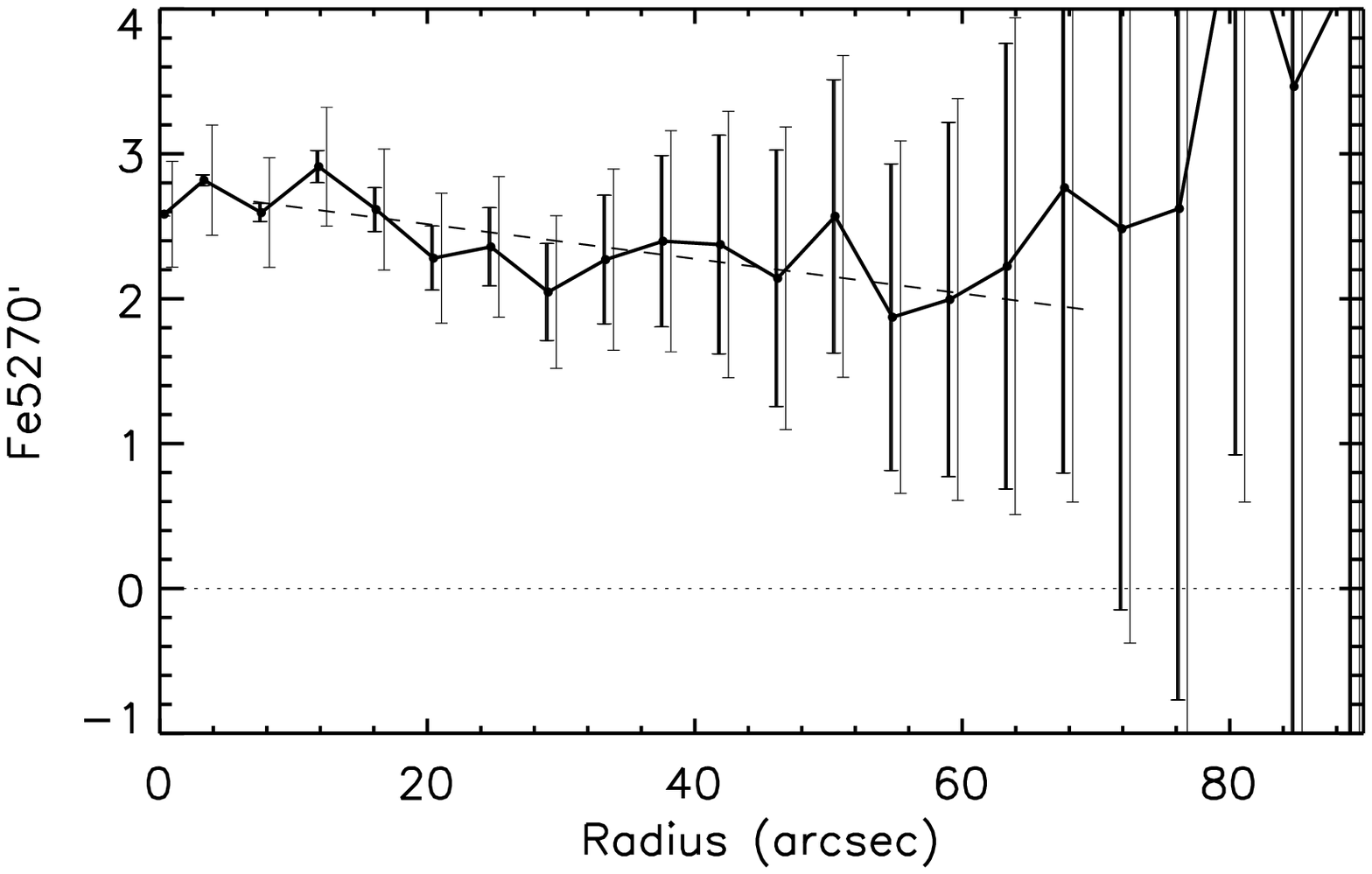}
    }
  }
  \vspace{9pt}
  \caption{Same as Fig.~\ref{5968_fig} but for NGC 6753}
  \label{6753_fig}
\end{figure*}

\begin{figure*}
\centering \vspace{30pt}
  \hbox{\hspace{0.5in} (a) \hspace{3.4in} (c)} 
  \centerline{\hbox{ \hspace{0.8in} 
\includegraphics[width=6cm]{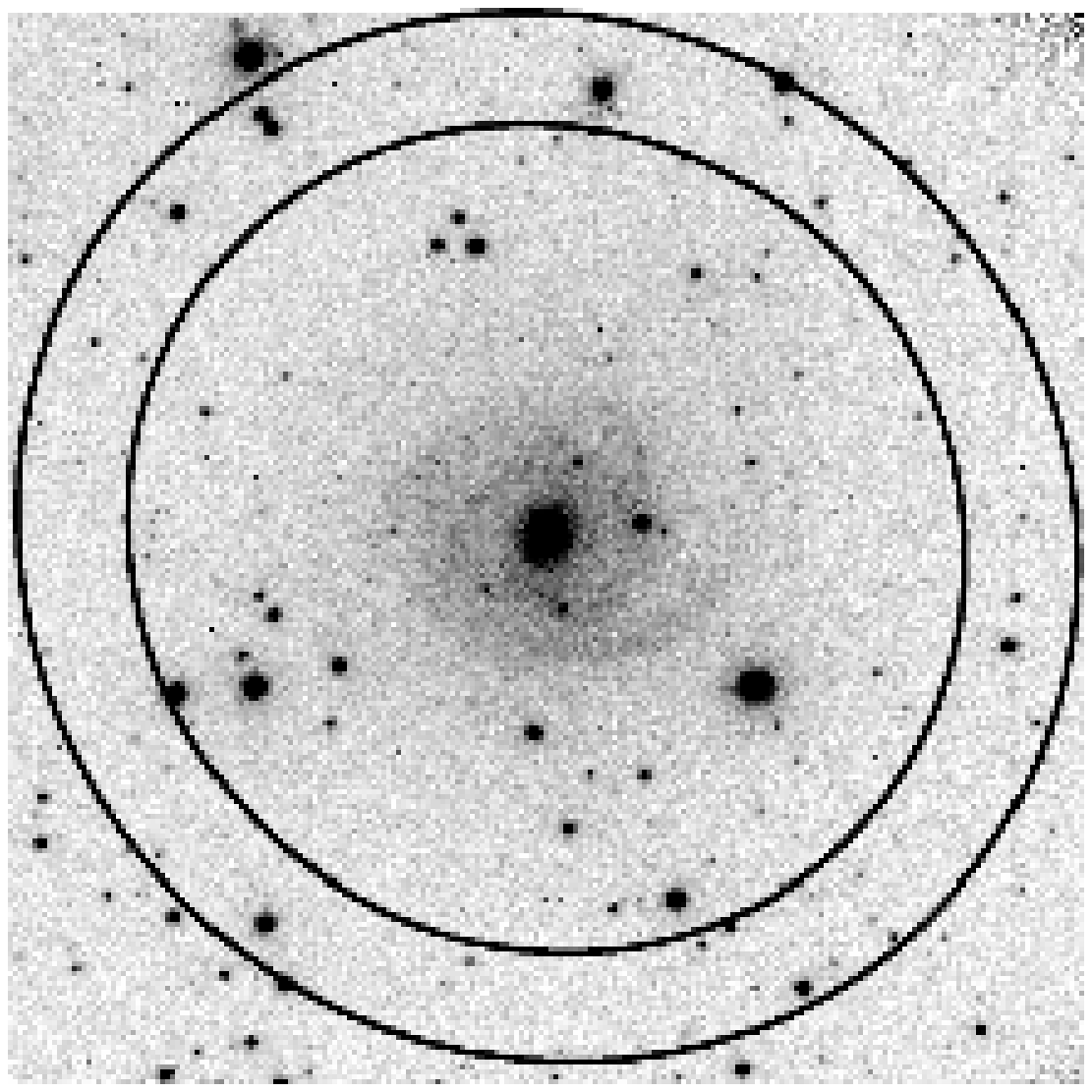}
    \hspace{0.3in}
\includegraphics[width=9cm]{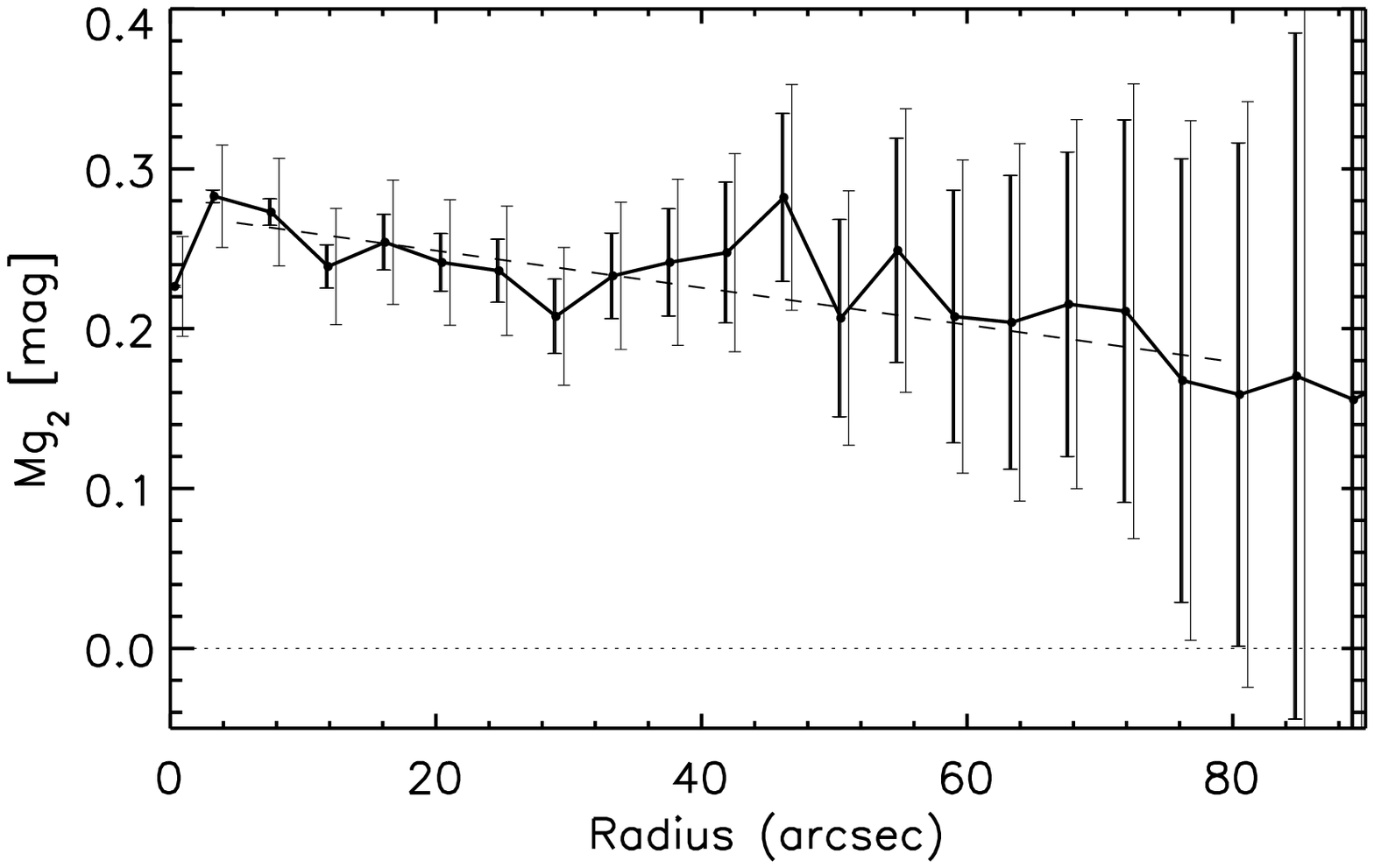}
    }
  }
  \vspace{30pt}
  \hbox{\hspace{0.5in} (b) \hspace{3.4in} (d)} 
  \centerline{\hbox{
\includegraphics[width=9cm]{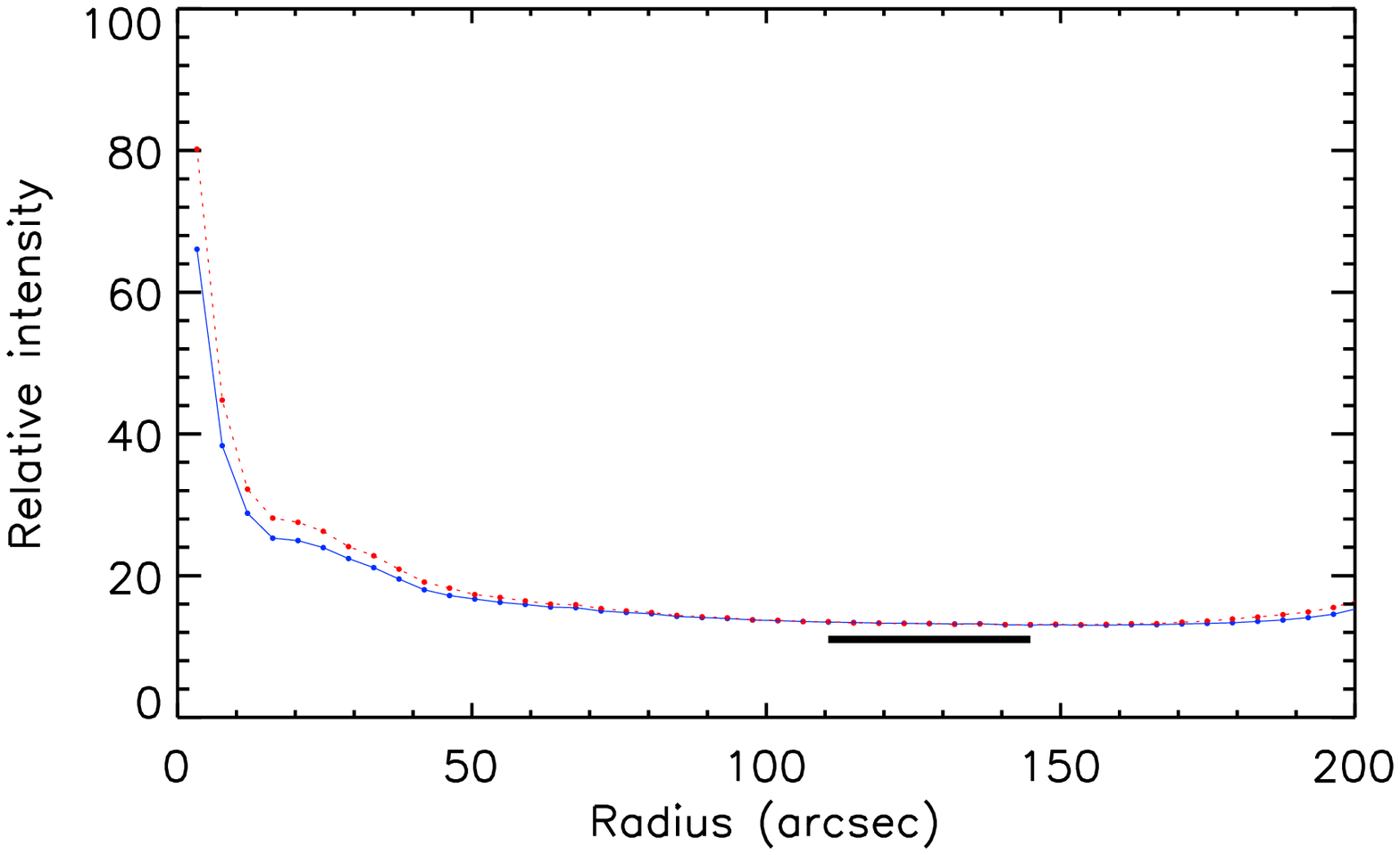}
\includegraphics[width=9cm]{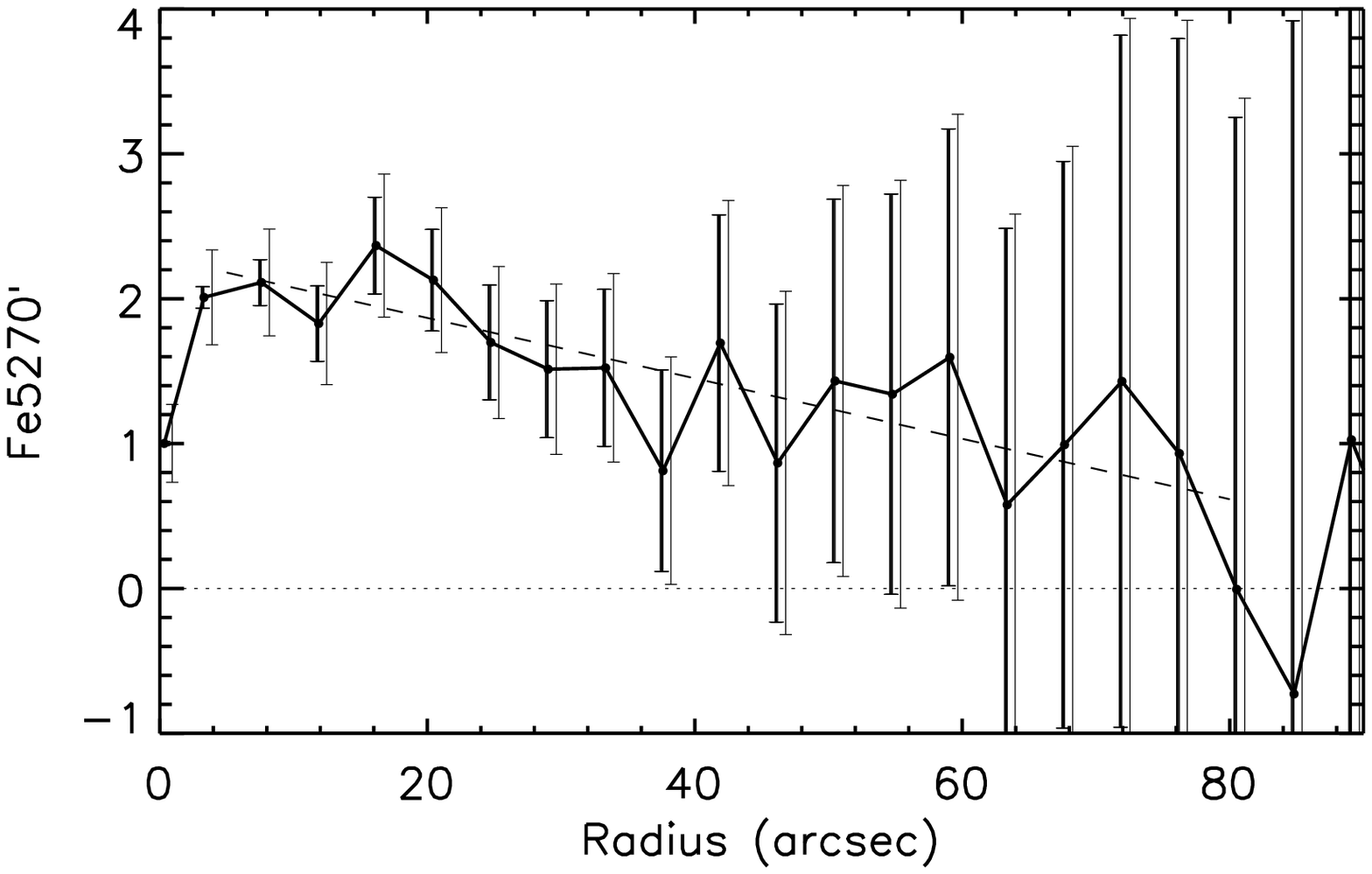}
    }
  }
  \vspace{9pt}
  \caption{Same as Fig.~\ref{5968_fig} but for NGC 6814}
  \label{6814_fig}
\end{figure*}

\begin{figure*}
\centering \vspace{30pt}
  \hbox{\hspace{0.5in} (a) \hspace{3.4in} (c)} 
  \centerline{\hbox{ \hspace{0.8in} 
\includegraphics[width=6cm]{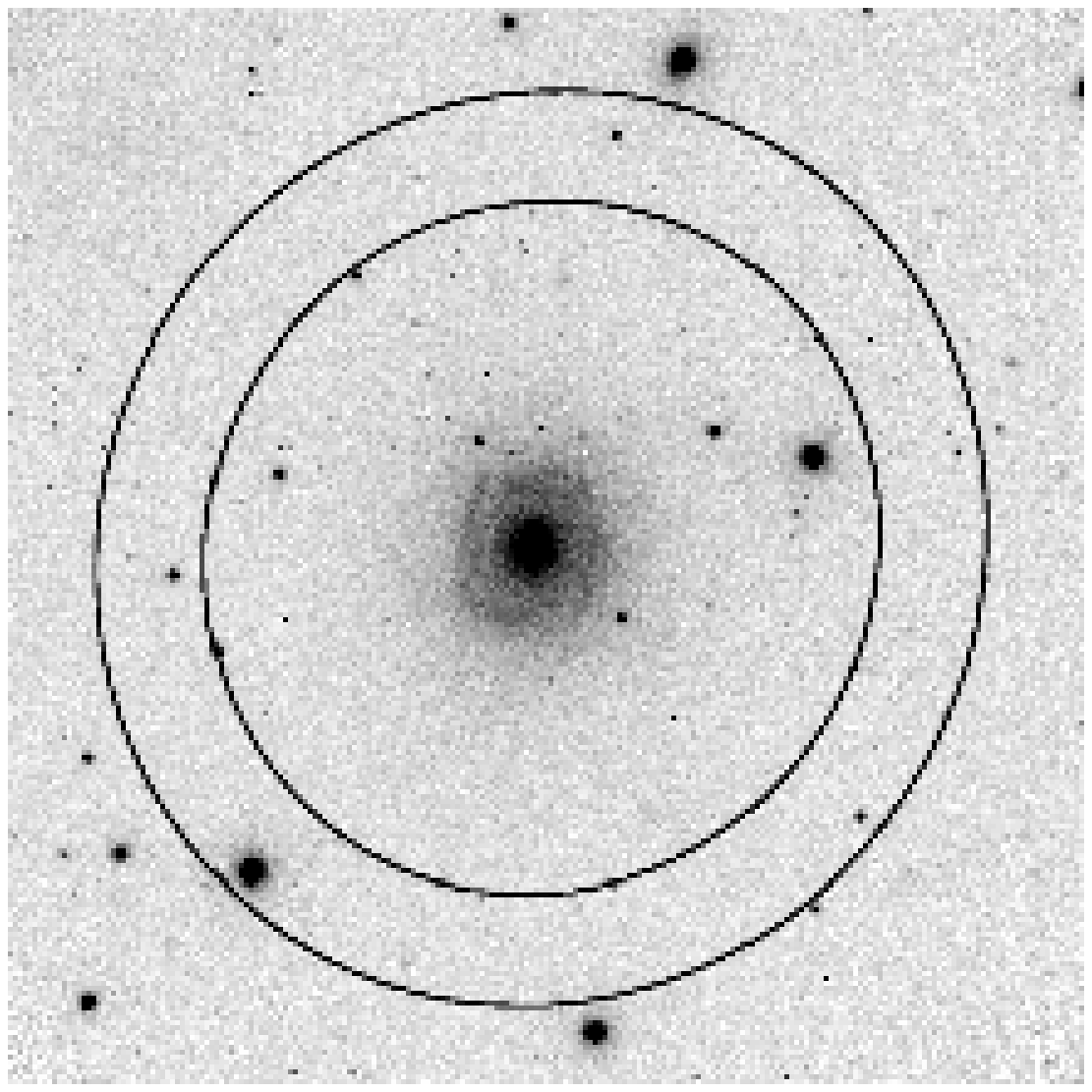}
    \hspace{0.3in}
\includegraphics[width=9cm]{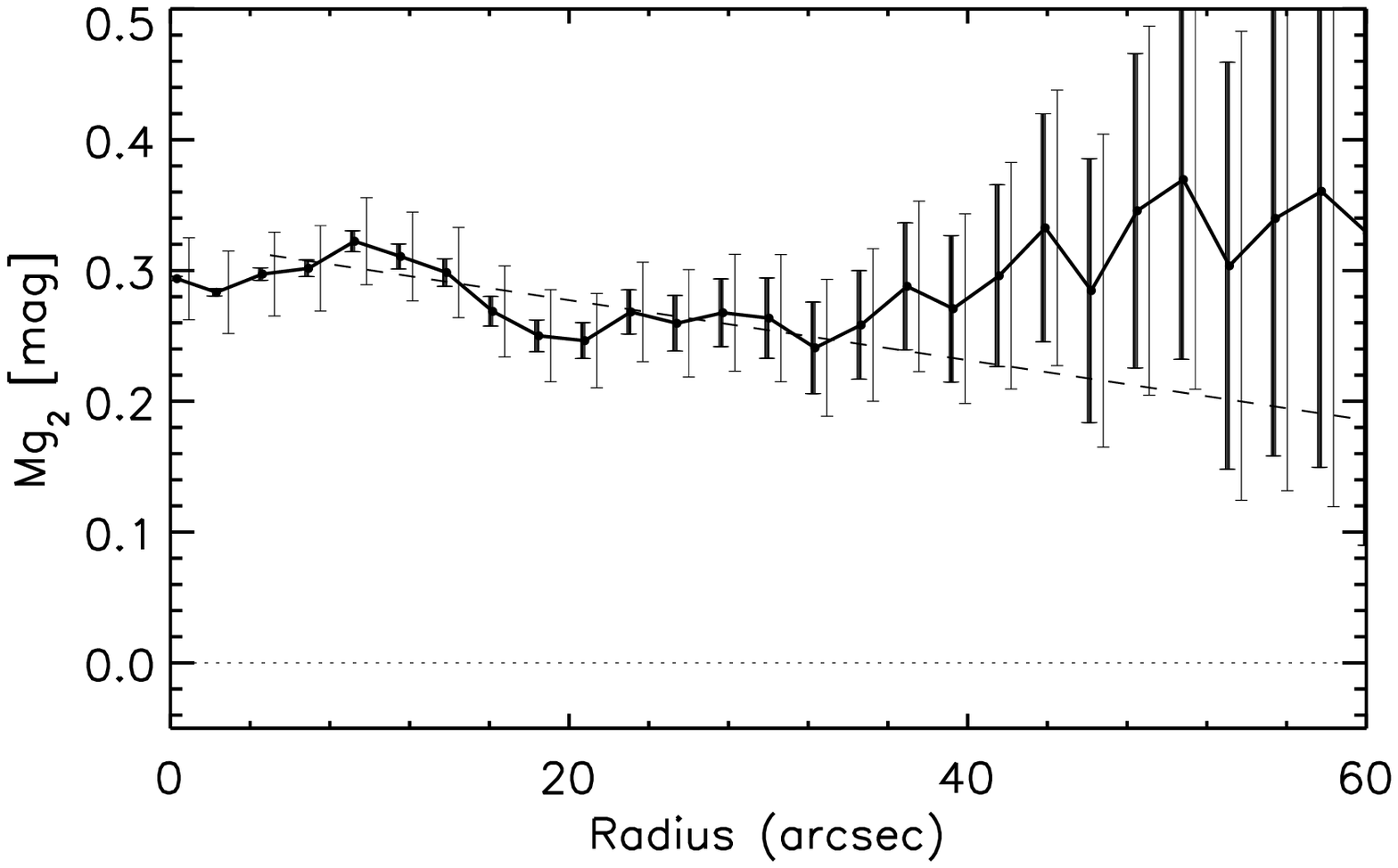}
    }
  }
  \vspace{30pt}
  \hbox{\hspace{0.5in} (b) \hspace{3.4in} (d)} 
  \centerline{\hbox{
\includegraphics[width=9cm]{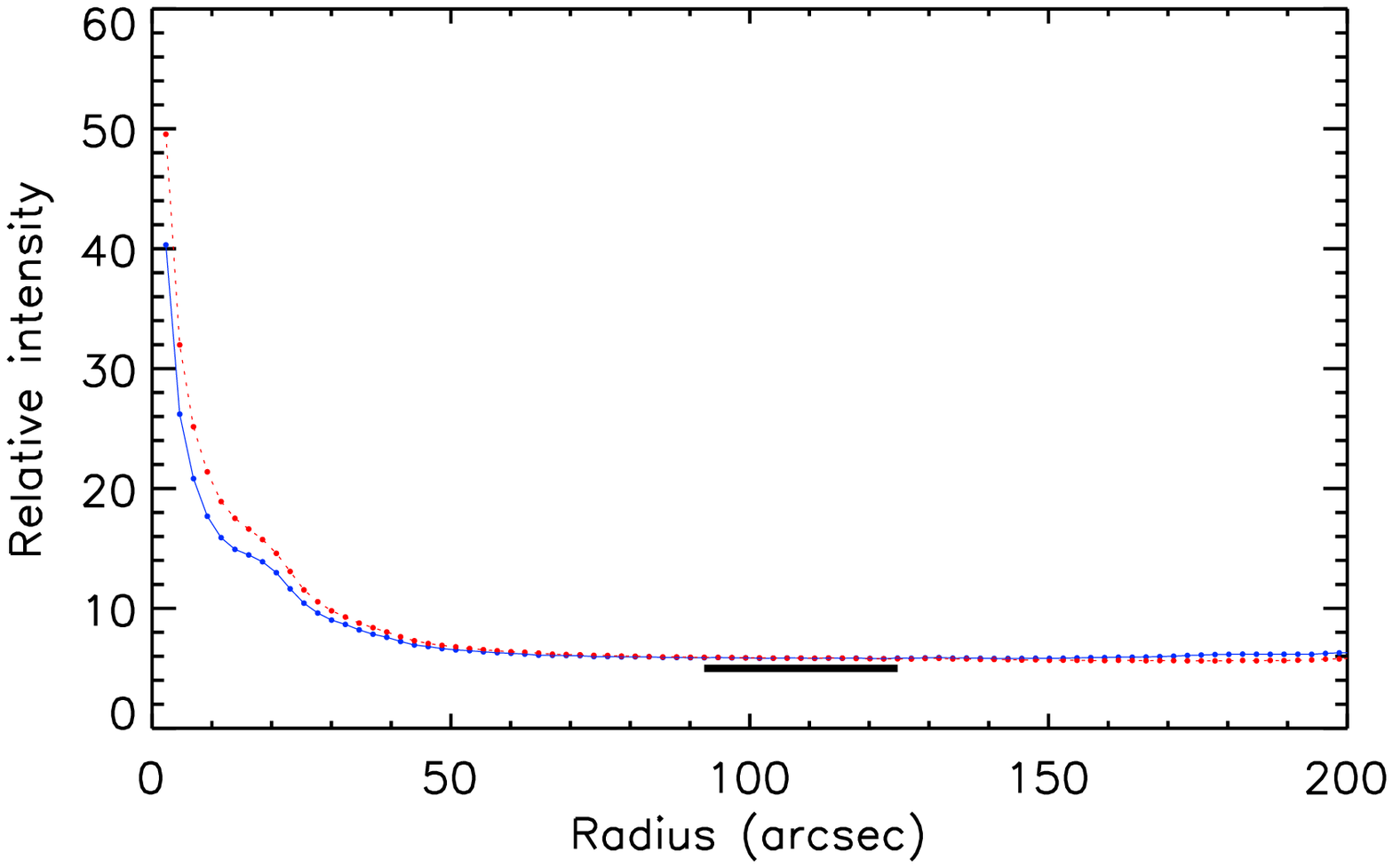}
\includegraphics[width=9cm]{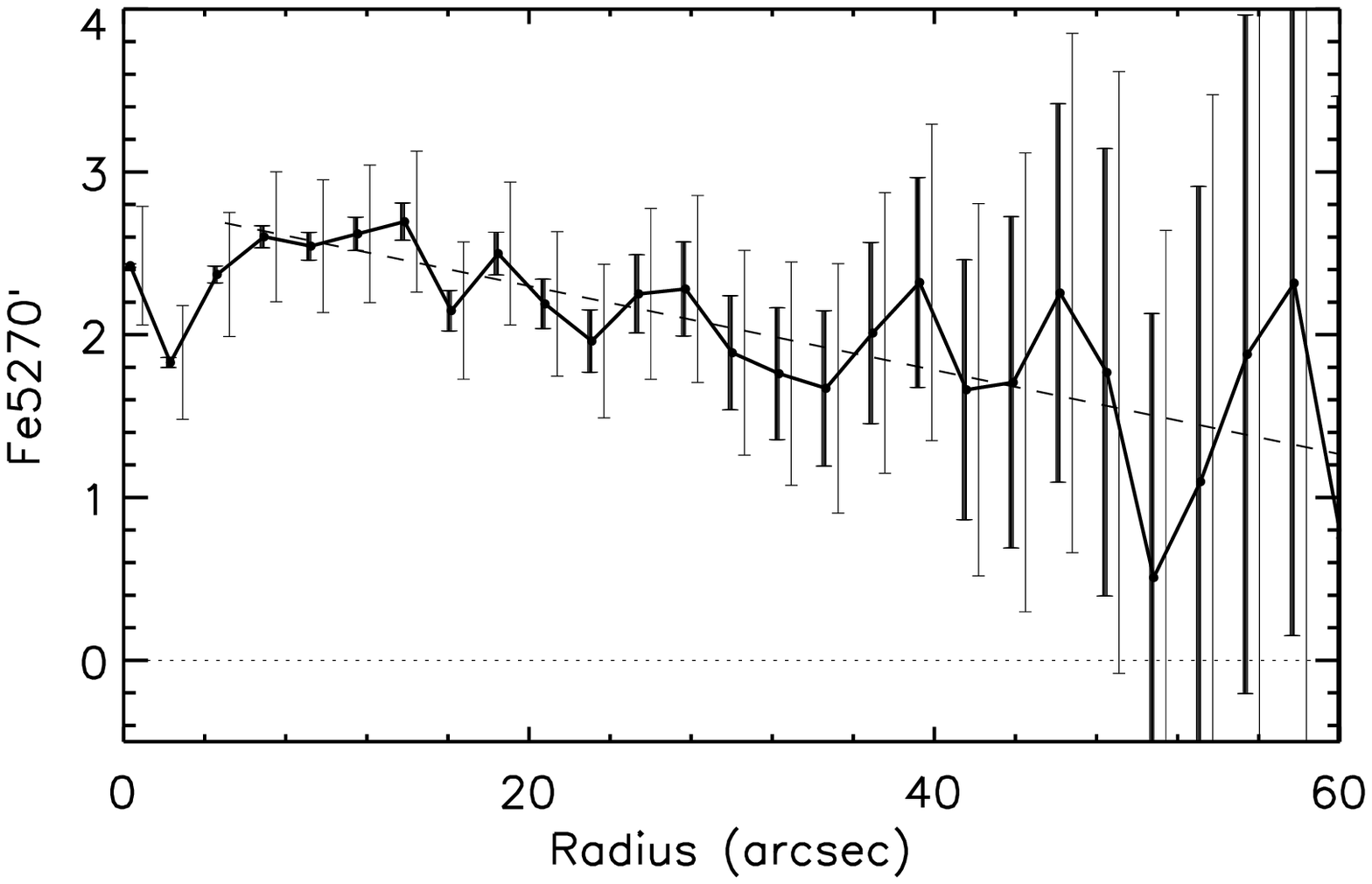}
    }
  }
  \vspace{9pt}
  \caption{Same as Fig.~\ref{5968_fig} but for NGC 6935}
  \label{6935_fig}
\end{figure*}

\begin{figure*}
\centering \vspace{30pt}
  \hbox{\hspace{0.5in} (a) \hspace{3.4in} (c)} 
  \centerline{\hbox{ \hspace{0.8in} 
\includegraphics[width=6cm]{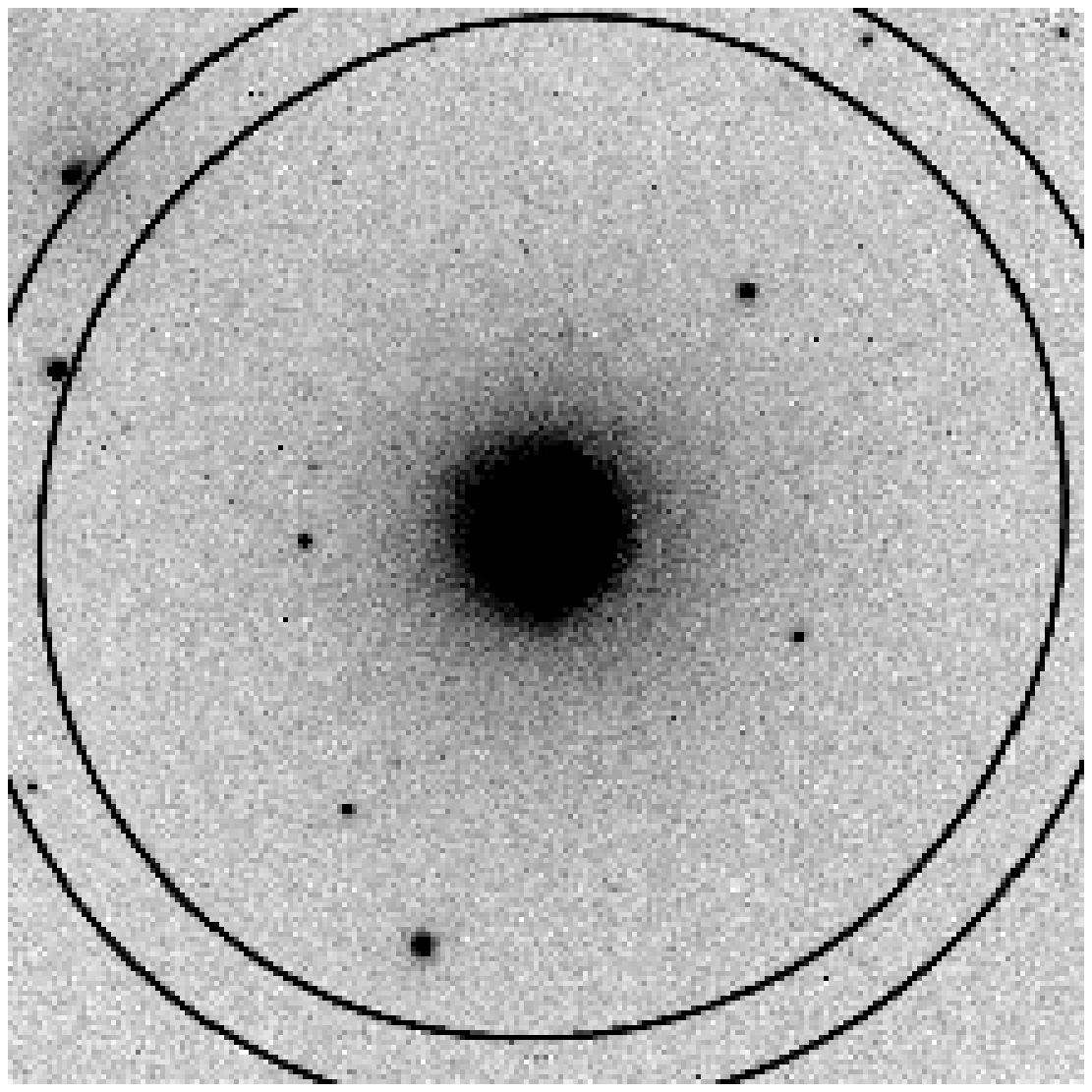}
    \hspace{0.3in}
\includegraphics[width=9cm]{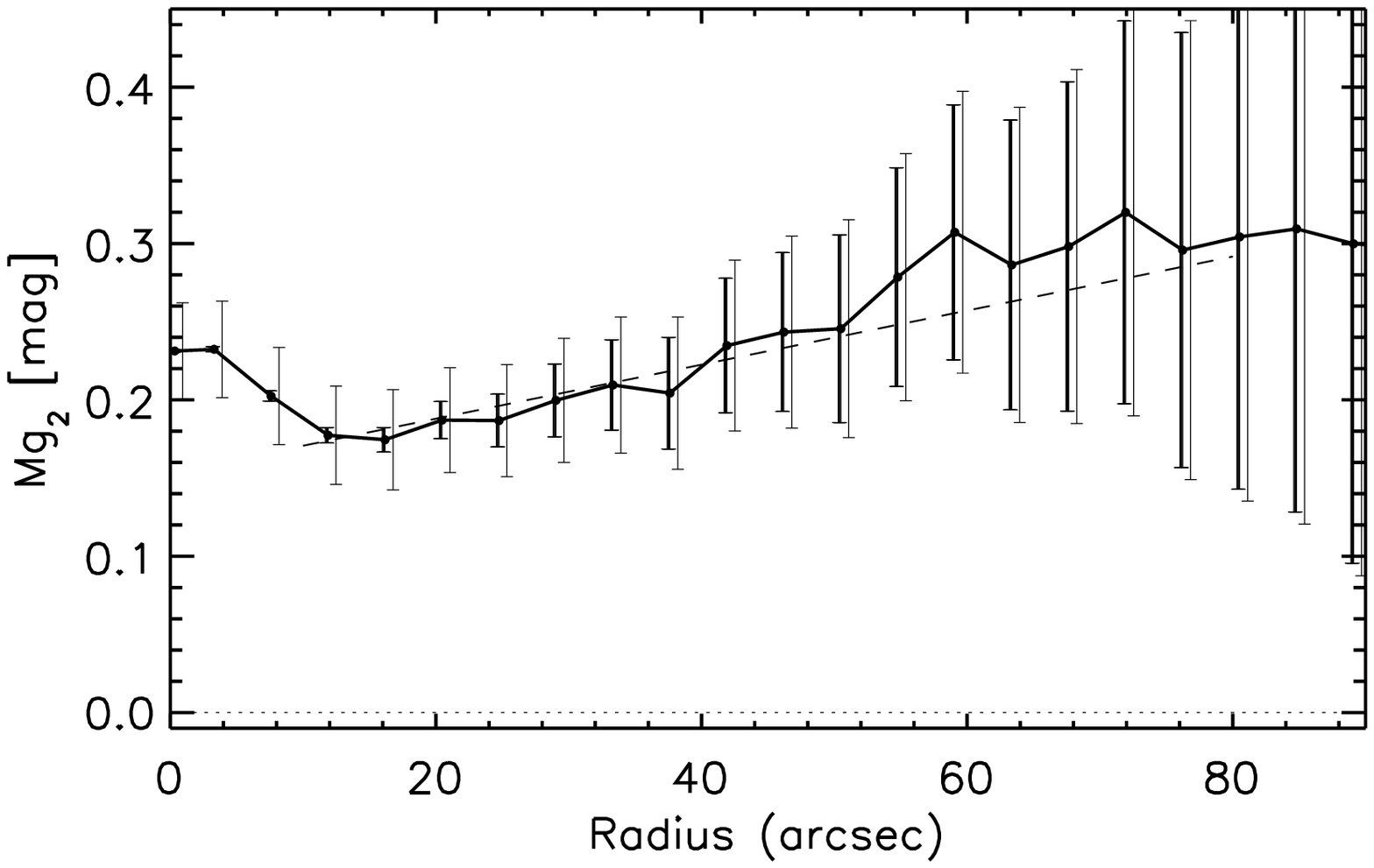}
    }
  }
  \vspace{30pt}
  \hbox{\hspace{0.5in} (b) \hspace{3.4in} (d)} 
  \centerline{\hbox{
\includegraphics[width=9cm]{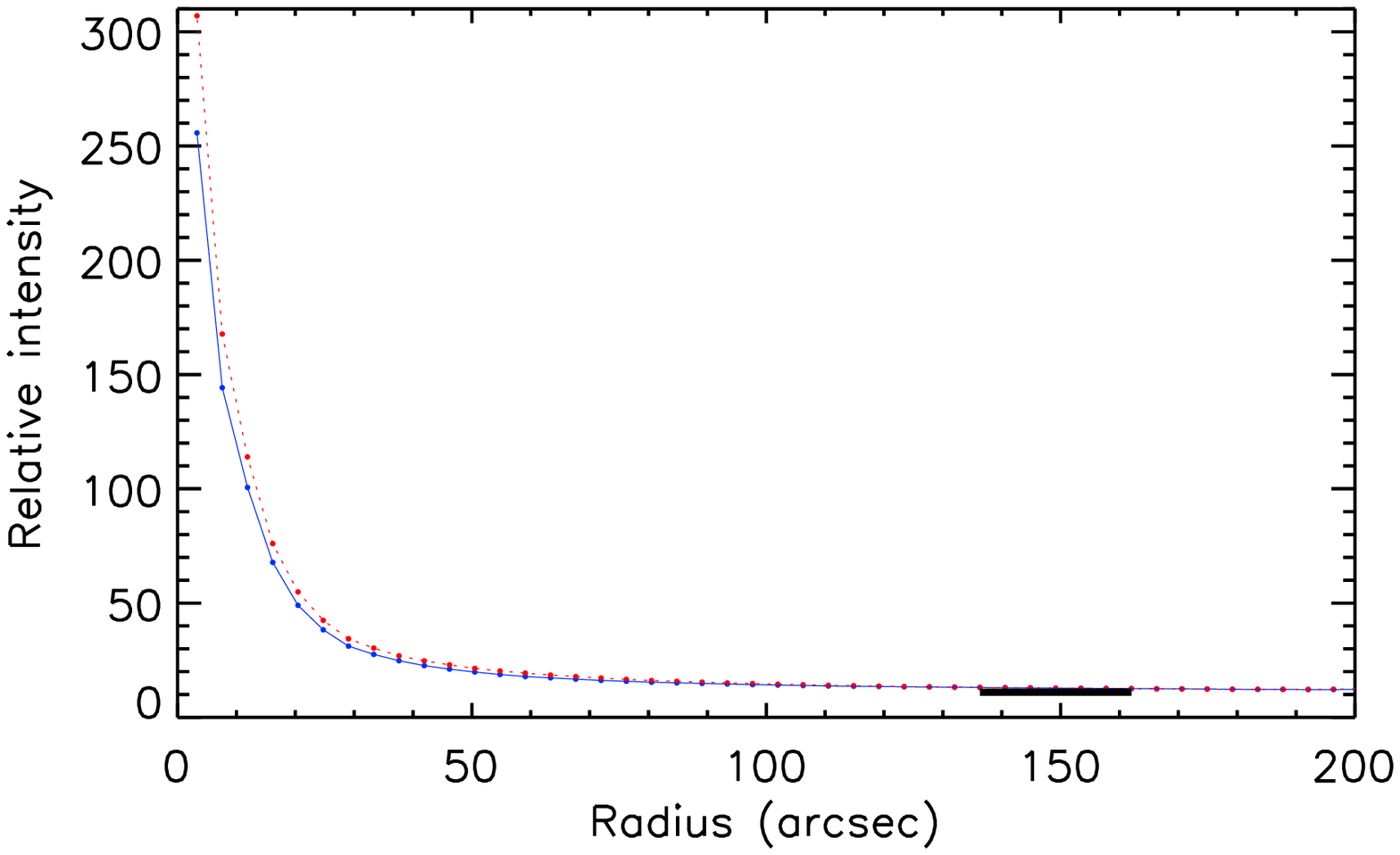}
\includegraphics[width=9cm]{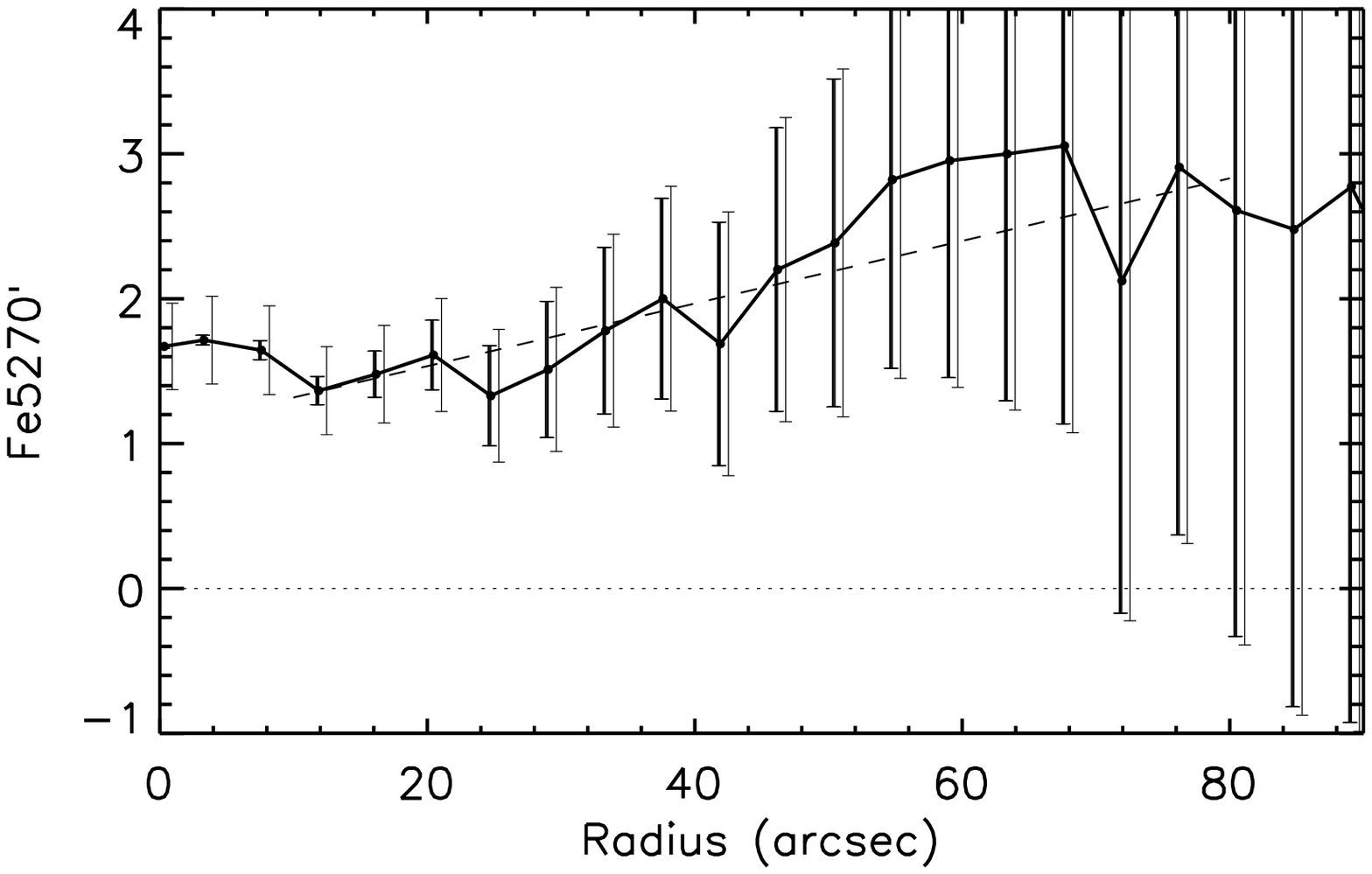}
    }
  }
  \vspace{9pt}
  \caption{Same as Fig.~\ref{5968_fig} but for NGC 7213}
  \label{7213_fig}
\end{figure*}

\begin{figure*}
\centering \vspace{30pt}
  \hbox{\hspace{0.5in} (a) \hspace{3.4in} (c)} 
  \centerline{\hbox{ \hspace{0.8in} 
\includegraphics[width=6cm]{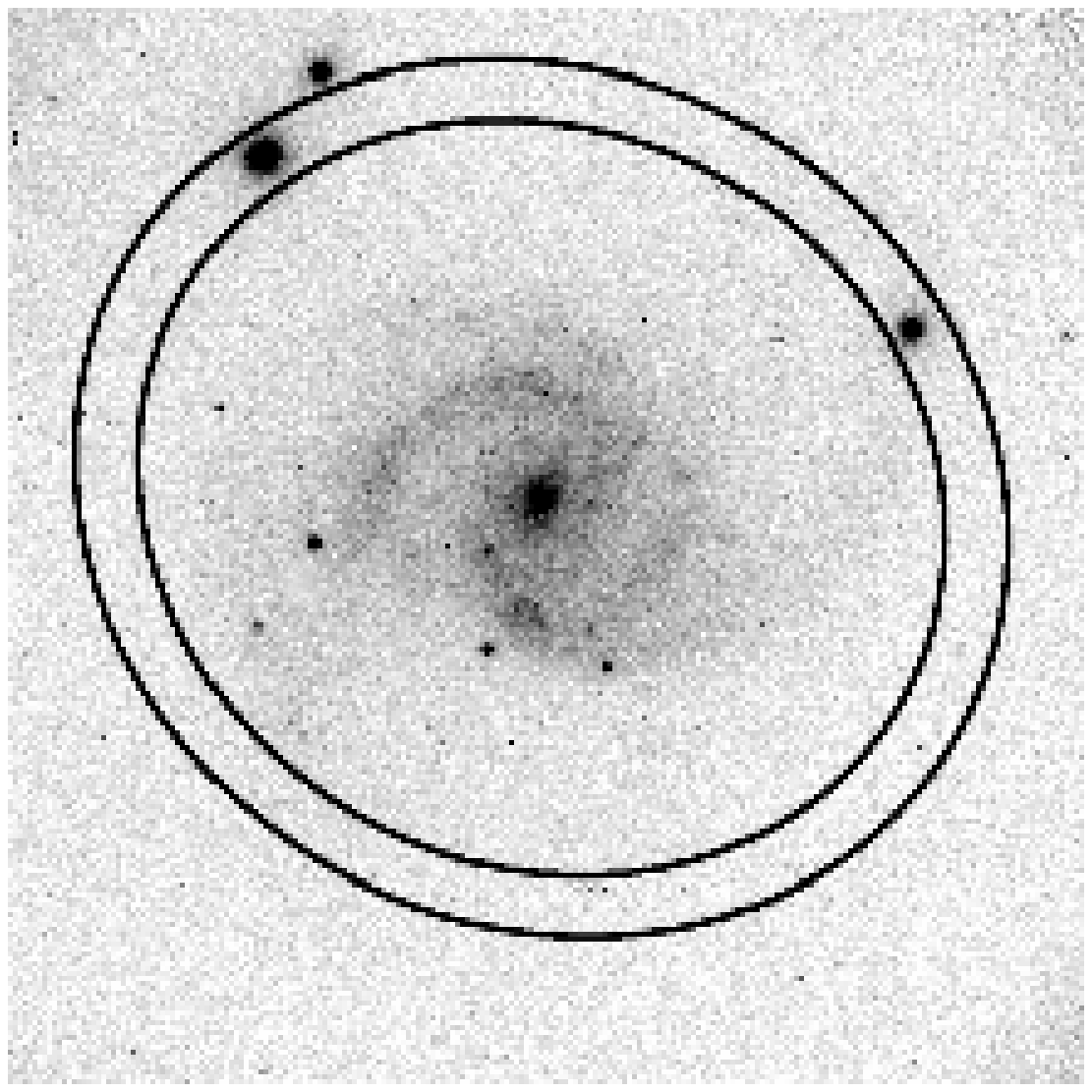}
    \hspace{0.3in}
\includegraphics[width=9cm]{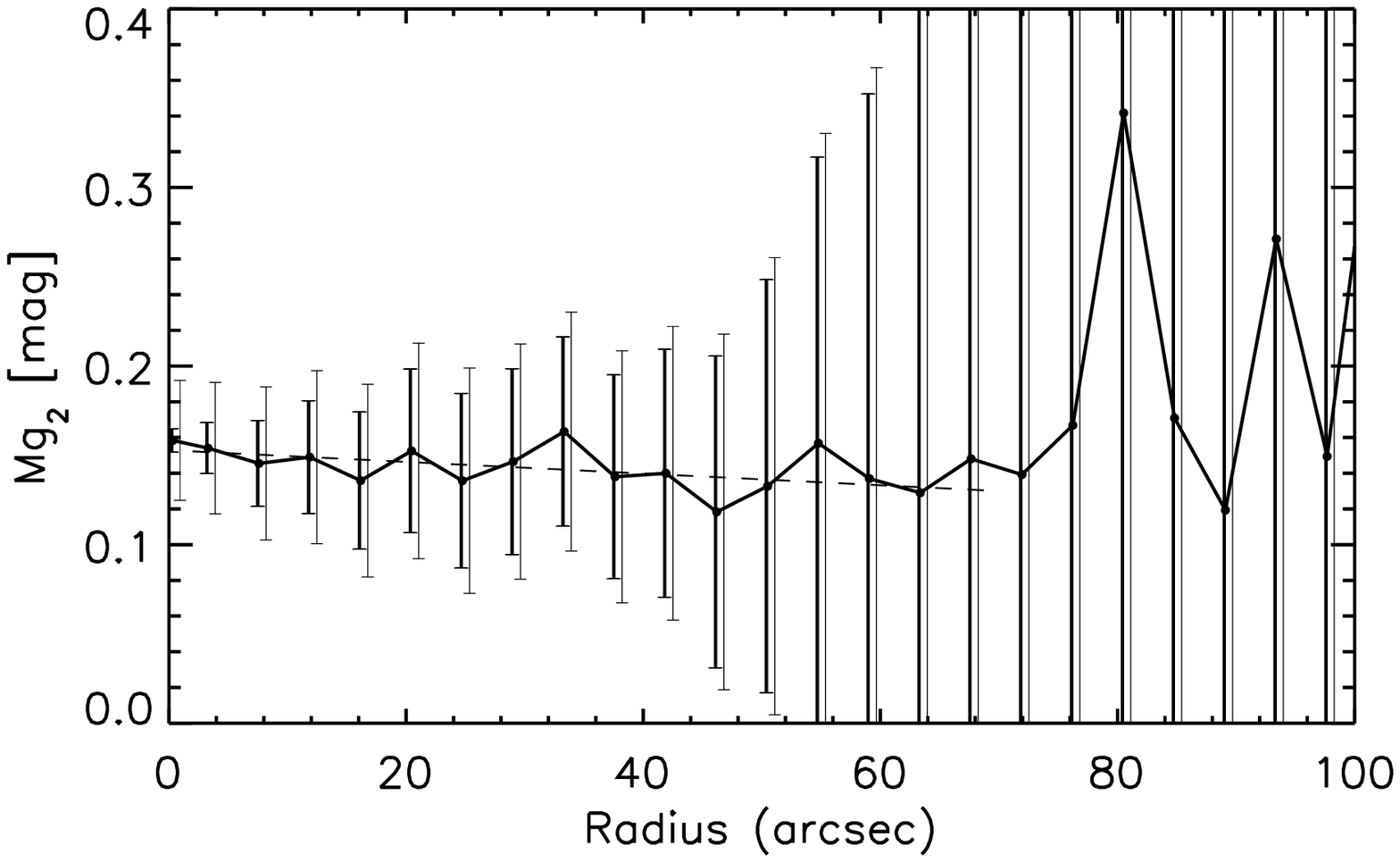}
    }
  }
  \vspace{30pt}
  \hbox{\hspace{0.5in} (b) \hspace{3.4in} (d)} 
  \centerline{\hbox{
\includegraphics[width=9cm]{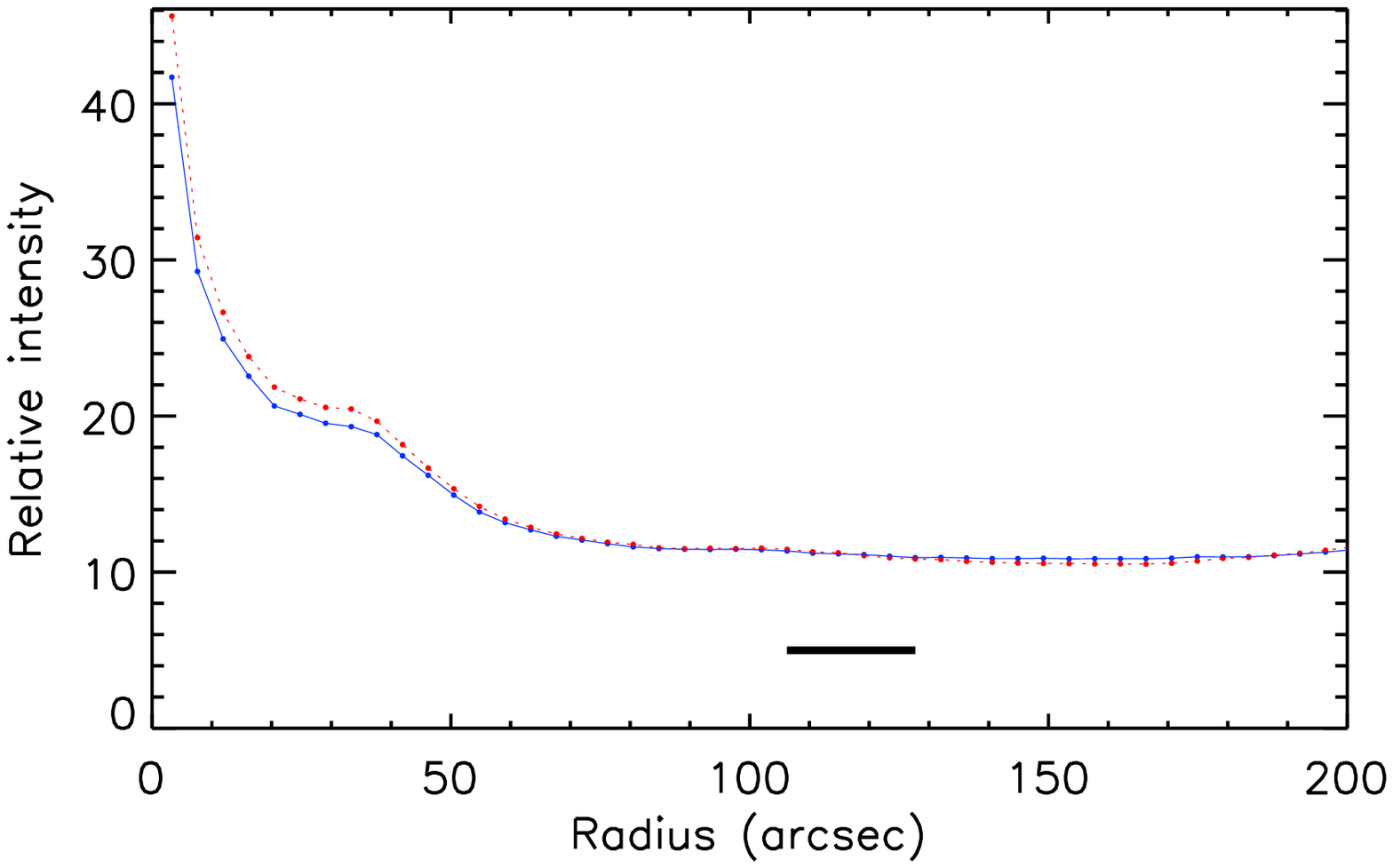}
\includegraphics[width=9cm]{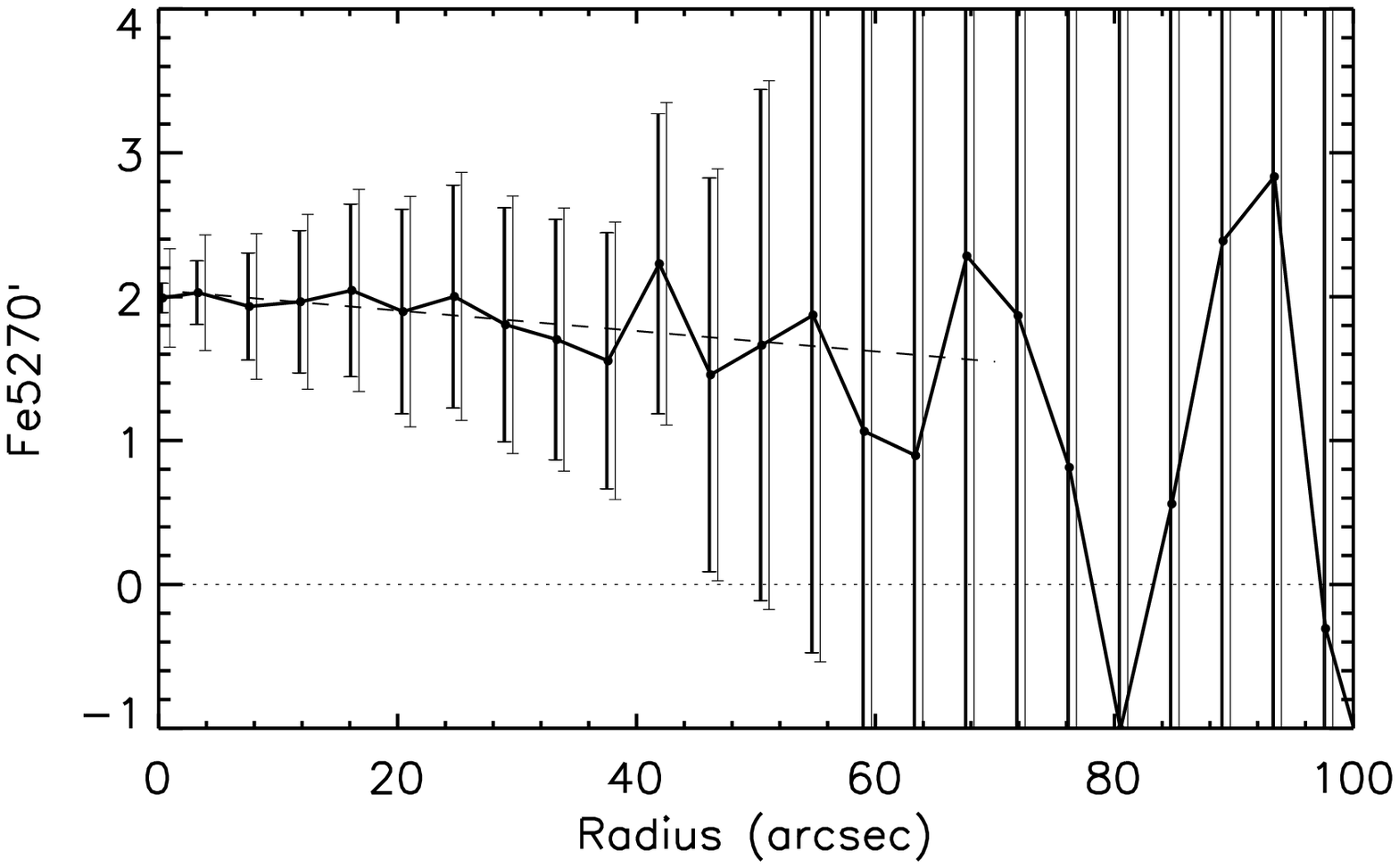}
    }
  }
  \vspace{9pt}
  \caption{Same as Fig.~\ref{5968_fig} but for NGC 7412}
  \label{7412_fig}
\end{figure*}

This galaxy has a small bright nucleus, a short bar, thin knotty arms,
and the lowest mean surface brightness of the galaxies in our sample
(see Table~\ref{gal_table}). The same overall features are seen in
both the {\mgtwo} and Fe5270 profiles (Fig.~\ref{5968_fig}), however
the {\mgtwo} gradient is steeper than that of Fe5270, giving NGC~5968
the highest ratio of $\Delta$Mg$_{2}$/$\Delta$Fe5270 in our sample.

\subsection{NGC 6221 - SB(s)bc pec}\label{ngc6221} 

This galaxy exhibits peculiar morphology and asymmetric arms due to
interaction with its neighbour NGC~6215 and several smaller dwarf
galaxies \citep{KD04}. A thick bar is evident and most of the current
star formation is offset from the nucleus. Cid Fernandes et~al. (1998)
measured the Mg~I+Mg~H EW out to a radius of 12 arcsec in NGC~6221.
Their EW peaked at about 5 arcsec before becoming diluted by about
50\% toward the nucleus (see their fig.~43). Our {\mgtwo} profile
peaks at a similar radius, but shows only moderate diminishment
towards the nucleus. This disagreement may be due to the different
methods used for continuum determination. The cause of the dilution of
this absorption feature is likely to be the presence of an active
galactic nucleus (AGN).  NGC~6221, NGC~6814 and NGC~7213 are all
classified as Seyfert galaxies and therefore produce non-thermal
continuum radiation and broad emission lines in the nucleus. The
superposition of an AGN continuum on top of the stellar spectrum can
act to decrease the measured strength of the absorption features in
the galaxy's core.

Curiously, the Fe5270 feature dips between about 5\,$-$\,10 arcsec and
rises toward the centre.  The {\mgtwo} line-strength shows no radial
variation in the disk, while the Fe5270 feature decreases fairly
steeply, such that NGC~6221 has the lowest ratio of
$\Delta$Mg$_{2}$/$\Delta$Fe5270 of the galaxies in our sample.

\subsection{NGC 6753 - (R')SA(r)b}\label{ngc6753} 

NGC~6753 has a nuclear ring-lens surrounded by a pseudoring, an
intermediate spiral region, and finally an outer pseudoring
\citep{CBB96}.  The arms are thin and have only small to moderate
current star formation rate  \citep{SB94}. We find that {\mgtwo}
increases steeply for this early-type galaxy in the inner 10
arcsec. Eskridge et al. (2003) found a similar rise in the (NUV --
$I_{814}$) colour over this radial range. We find evidence for a bump
in {\mgtwo} at about 45 arcsec that seems to correspond to an outer ring of
ongoing star formation, which contains many \mbox{H\,{\sc ii}} regions. Neither the
{\mgtwo} bump at 45 arcsec, nor the change in slope at 10 arcsec, is reflected
in the Fe5270 profile.

\subsection{NGC 6814 - SAB(rs)bc}\label{ngc6814} 

This galaxy has a thick bar and a grand design spiral structure
(M{\'a}rquez et al. 1999), whose spiral arms contain many
\mbox{H\,{\sc ii}} regions (Gonzalez Delgado et al. 1997).
NGC~6814 is another Seyfert galaxy observed by Cid Fernandes et
al. (1998). They found that the Mg~I+Mg~H feature was diluted by 22\%
in the nuclei, in agreement with a $\sim$~20\% decrease in the
{\mgtwo} EW we observe between 5 and 0 arcsec. The Fe5270
line-strength also appears significantly diluted in the nuclear
region. The line strengths of both spectral features show moderate
declines with radius.

\subsection{NGC 6935 - (R)SA(r)a}\label{ngc6935} 

This galaxy is highly circular with a boxy inner H$\alpha$ ring,
tightly wound spiral arms, and a small bright nucleus (Crocker, Baugus
\& Buta 1996). It is separated from NGC~6937 by 246 arcsec (Sandage \&
Bedke 1994) but the pair do not appear to be interacting. As mentioned
above, there is good agreement between our central Fe5270 value and
that from Idiart et al. (1996), whereas we obtain a slightly higher
value for {\mgtwo}. The strengths of both indices show clear radial
declines in the galactic disk and maxima at around 10 arcsec. Both
indices decrease in the inner 10 arcsec, while Fe5270 drops
significantly at a radius of few arcsec.

\subsection{NGC 7213 - SA(s)0 LINER}\label{ngc7213} 

This well-studied LINER/Seyfert galaxy has the highest surface
brightness and earliest Hubble-type in our sample. NGC~7213 is also
unique among our galaxies in having \emph{positive} {\mgtwo} and
Fe5270 gradients in the disk.

Interestingly, Cid Fernandes et al. (1998) found no dilution of the
Mg~I+Mg~H EW in the heart of NGC~7213, but strong dilution of features
at shorter wavelengths. The EW of their Mg feature increases by about
50\% from 25 arcsec to the nucleus. We found that our {\mgtwo} ebbs at a
radius of $\sim$15 arcsec before increasing by $\sim$ 35\% (in EW)
toward the core. Fe5270 shows much less pronounced enhancement toward
the core but a comparable steep positive gradient in the outer disk.
The positive gradients may be associated with a ring of \mbox{H\,{\sc
ii}} regions located at 20\,$-$\,40 arcsec in radius (Evans et
al. 1996).

Corsini et al. (2003) note that the gaseous and stellar disks of
NGC~7213 are decoupled within 40 arcsec. They point out that the
\mbox{H\,{\sc ii}} circumnuclear ring is not coincident with the
stellar disk identified by Mulchaey, Regan \& Kundu
(1997). \mbox{H\,{\sc i}} and H$\alpha$ measurements reveal that
NGC~7213 is a disturbed system undergoing a merger process that has
warped its disk (Hameed et al. 2001) and may be responsible for the
gaseous/stellar decoupling.

\subsection{NGC 7412 - SAB(s)c}\label{ngc7412} 

The radial line-strength gradients for NGC~7412 are fairly flat and
featureless: possible evidence of smoothing due to the presence of the
bar with length 30 arcsec (Saraiva Schroeder et al. 1994) . Photometry
results from these authors show that the $V-I$ and $V-R$ colours are
almost constant along the bar.  Unfortunately, our estimated gradients
are particularly uncertain for this galaxy due to large uncertainties
in the background sky level.

\subsection{NGC 7637 - SA(r)bc}\label{ngc7637} 

\begin{figure*}
\centering \vspace{30pt}
  \hbox{\hspace{0.5in} (a) \hspace{3.4in} (c)} 
  \centerline{\hbox{ \hspace{0.8in} 
\includegraphics[width=6cm]{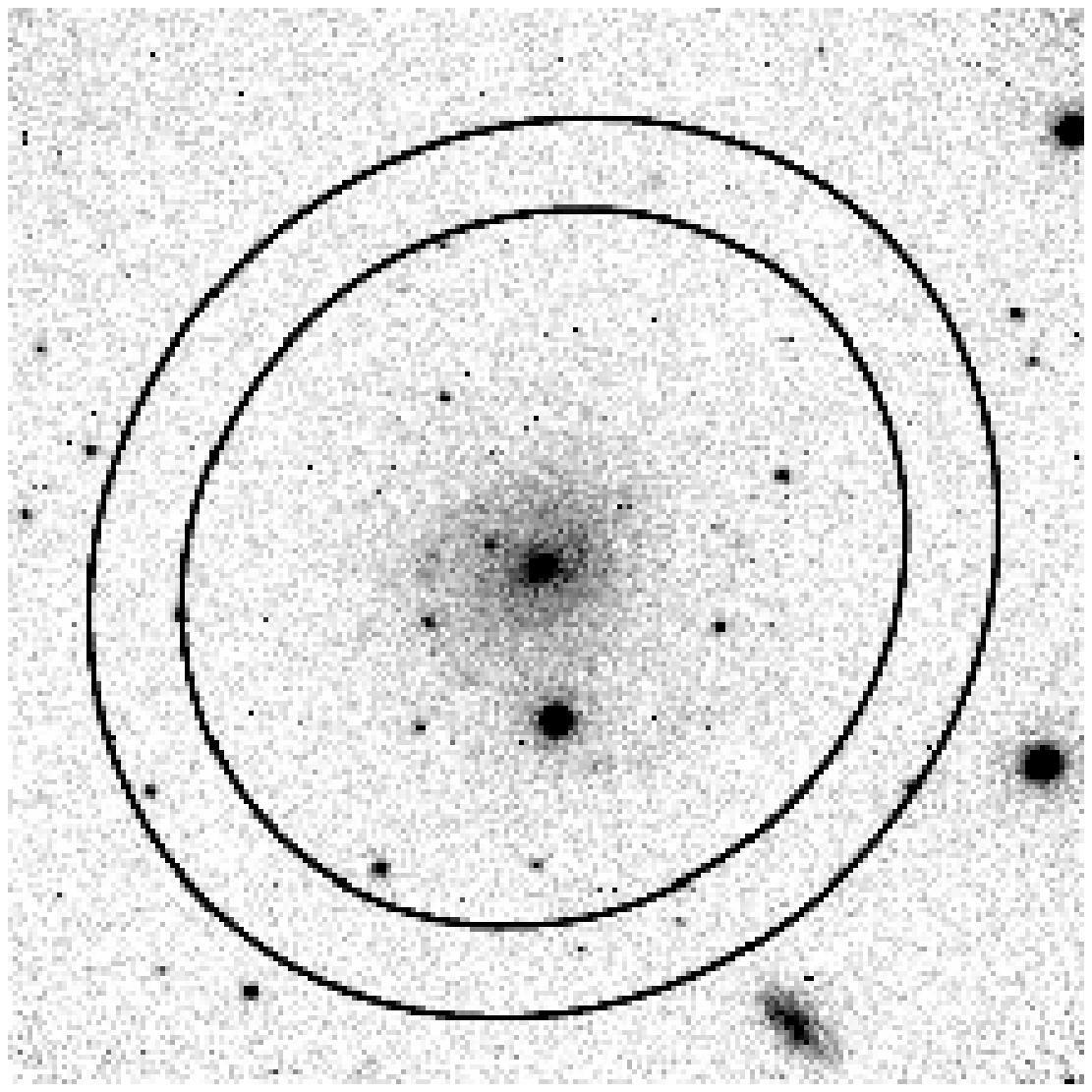}
    \hspace{0.3in}
\includegraphics[width=9cm]{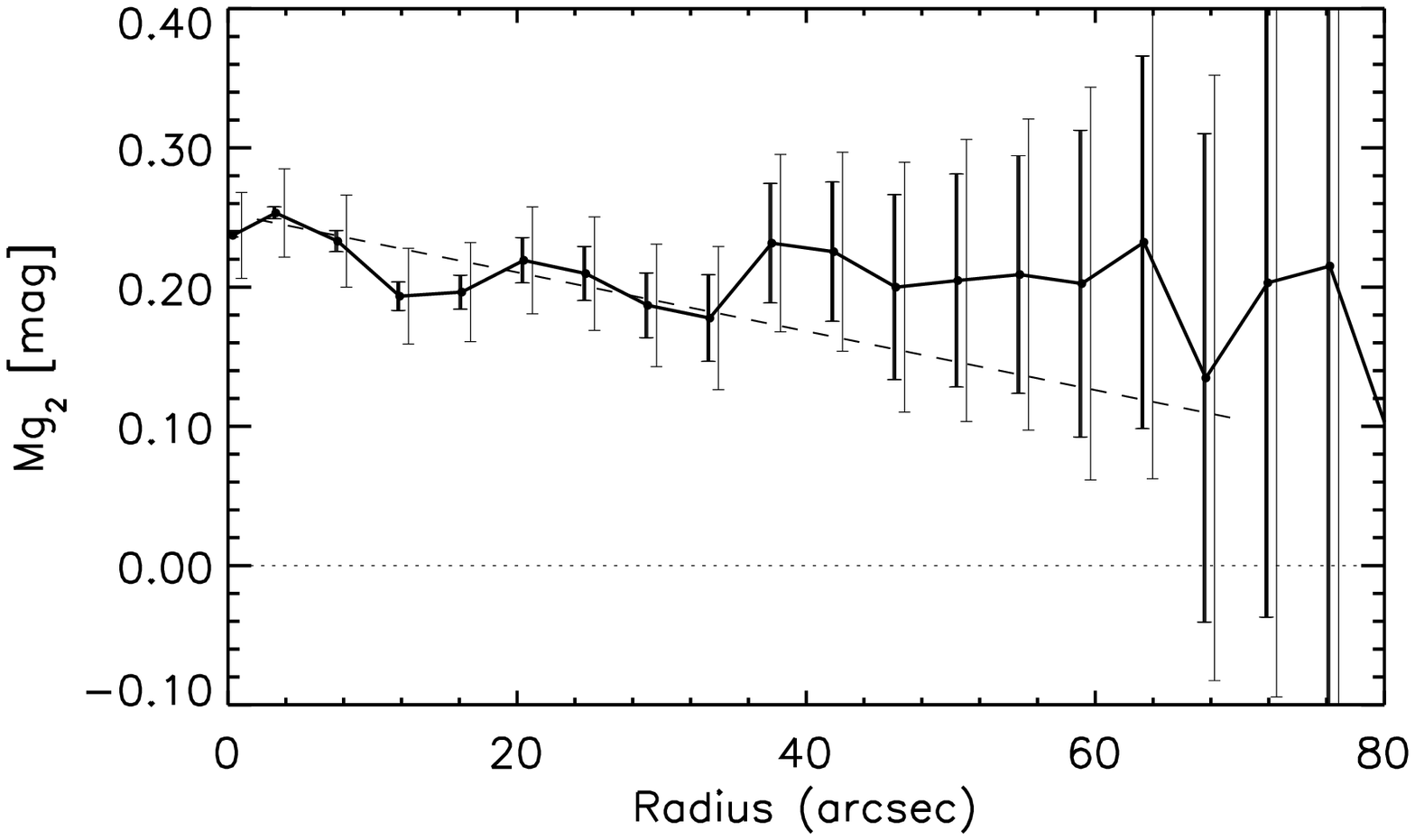}
    }
  }
  \vspace{30pt}
  \hbox{\hspace{0.5in} (b) \hspace{3.4in} (d)} 
  \centerline{\hbox{
\includegraphics[width=9cm]{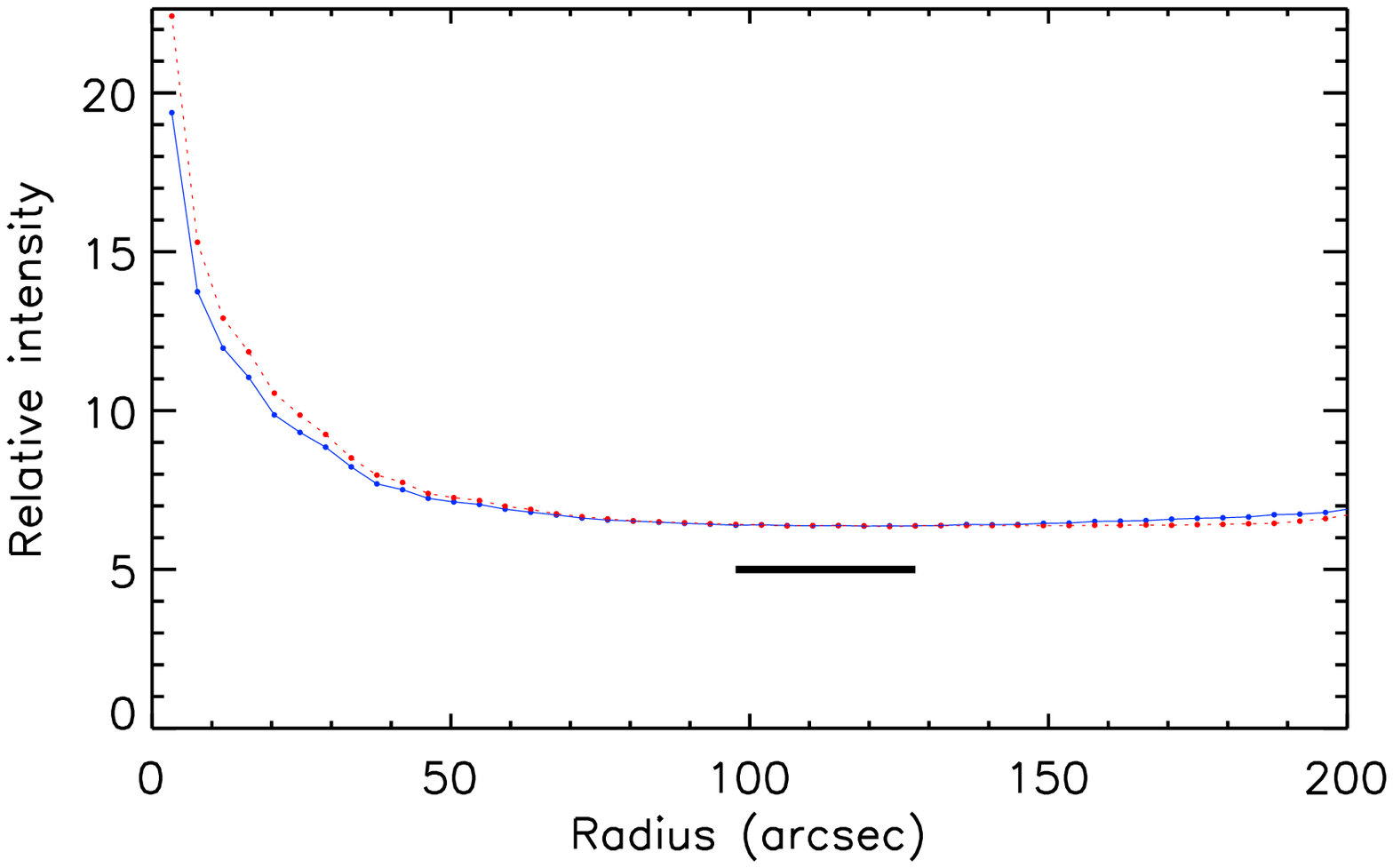}
\includegraphics[width=9cm]{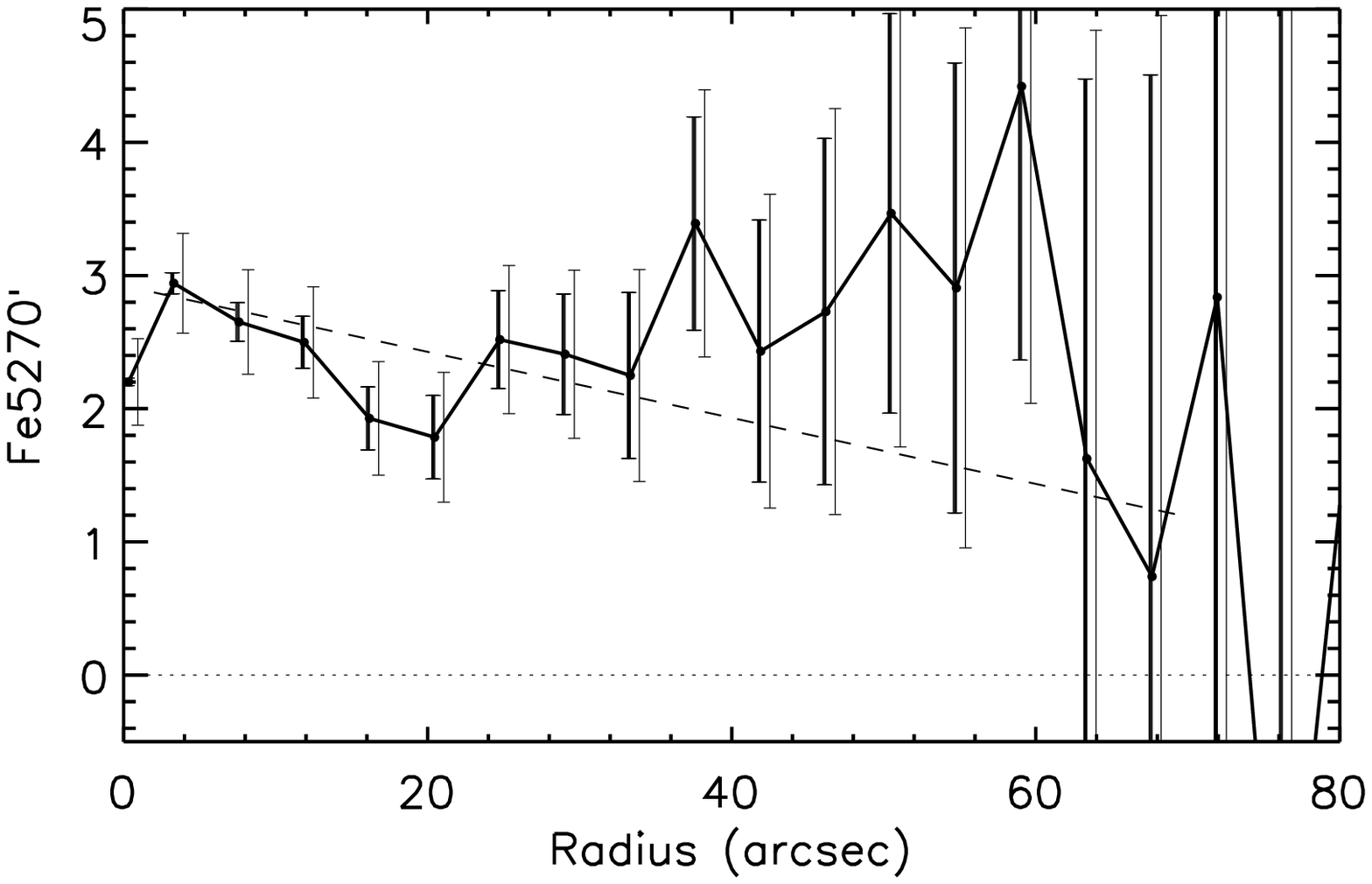}
    }
  }
  \vspace{9pt}
  \caption{Same as Fig.~\ref{5968_fig} but for NGC 7637}
  \label{7637_fig}
\end{figure*}

This late-type galaxy (Corwin, de Vaucouleurs \& de Vaucouleurs 1985)
shows considerable radial variation in {\mgtwo} and Fe5270. It is
interacting with a companion at a distance of about 2.4 arcmin
(evident in the corner of the image in Fig.~\ref{7637_fig}) that is
likely responsible for asymmetrical spiral arms and tidal features.
We observe dips in both spectral indices at the galaxy core as
expected for an AGN host, despite NGC~7637 not having a Seyfert
classification.  We note that Pastoriza, Donzelli \& Bonatto (1999)
have also postulated that this galaxy has an active nucleus, based on
their analysis of its emission spectrum. Our results support this
proposition.

The {\mgtwo} and Fe5270 features do not trace each other
perfectly. For instance, {\mgtwo} declines sharply from $\sim$\,5   to
15 arcsec in radius followed by a bump at about 20 arcsec, whereas Fe5270
continues to fall from $\sim$ 5 to 20 arcsec with a possible bump around
25\,$-$\,30 arcsec. Further investigation is required to interpret the apparent
offsets in the {\mgtwo} and Fe5270 behaviour as a function of radius.


\section{Summary}\label{discussion}

We have calibrated the technique of absorption line imaging with the
Anglo-Australian Telescope's Taurus Tunable Filter using Lick standard
stars and have demonstrated the sensitivity of this method to the
behaviour of {\mgtwo} and Fe5270 absorption features across the face
of disk galaxies. The variation of {\mgtwo} and Fe5270 line-strengths
as a function of radius is presented for a sample of eight galaxies.
The greatest source of uncertainty in our derived line-strength
gradients is associated with background sky subtraction.  As well as
yielding overall line-strength gradients, the detailed shapes of the
profiles may reveal phenomena including merger-induced star formation
(NGC~7213), \mbox{H\,{\sc ii}} rings (NGC~6753 and NGC~7213) and
galactic bars (NGC~7412). There is also evidence that NGC~7637 hosts
an active galactic nucleus. Conversion of the line-strengths presented
in this paper into absolute abundances involves stellar population
synthesis modelling, which will be the subject of the second paper in
this series. Ultimately, the inferred relative abundances of Mg and Fe
can be used to piece together star formation histories and galactic
chemical evolution.

Future generations of tunable filters on larger telescopes will be
more sensitive to stellar absorption features, enabling this type of
study to be extended to lower surface brightness galaxies. Moreover,
greater sensitivity will minimise the importance of azimuthal
averaging in order to achieve acceptable signal-to-noise. This will
facilitate comparisons between arm and interarm regions, thus
providing insight into the mechanism of spiral arm triggered star
formation.

\section*{Acknowledgements}

We wish to acknowledge our gratitude to Joss Bland-Hawthorn, for first
suggesting this avenue of research. We are grateful to the staff of
the Anglo-Australian Observatory for maintaining this unique facility,
to the Isaac Newton Group of Telescopes (La Palma) for the use of
their f126 blocking filter, and to the Australian Time Assignment
Committee (ATAC) for awarding us the time to prove this
technique. Financial support from the Australian Research Council
(ARC) is gratefully acknowledged. This research has made use of NASA's
Astrophysics Data System Bibliographic Services (ADS), the HyperLeda
database (http://leda.univ-lyon1.fr), as well as the NASA/IPAC
Extragalactic Database (NED) which is operated by the Jet Propulsion
Laboratory, California Institute of Technology, under contract with
the National Aeronautics and Space Administration. YF thanks the
Australian Federation of University Women (SA) for their support
through the Daphne Elliot Bursary.

\bsp

\label{lastpage}

\appendix

\section{Mg$_2$ and Fe5270 radial profiles for eight galaxies}\label{appendix}

\begin{table}
\caption{  NGC~5968}
\begin{tabular}{lcc} \hline
Radius & Mg$_2$ & Fe5270 \\ (arcsec) & (mag) & ($\AA$) \\
\hline
    0.0  &    0.223 $\pm$    0.031 &     3.22 $\pm$     0.38 \\
    3.3  &    0.229 $\pm$    0.031 &     3.40 $\pm$     0.41 \\
    7.59 &    0.253 $\pm$    0.033 &     3.52 $\pm$     0.45 \\
   11.88 &    0.234 $\pm$    0.035 &     3.04 $\pm$     0.47 \\
   16.17 &    0.229 $\pm$    0.037 &     2.82 $\pm$     0.49 \\
   20.46 &    0.225 $\pm$    0.039 &     2.83 $\pm$     0.52 \\
   24.75 &    0.177 $\pm$    0.042 &     2.29 $\pm$     0.58 \\
   29.04 &    0.184 $\pm$    0.044 &     2.71 $\pm$     0.64 \\
   33.33 &    0.189 $\pm$    0.050 &     2.42 $\pm$     0.73 \\
   37.62 &    0.167 $\pm$    0.055 &     2.43 $\pm$     0.85 \\
   41.91 &    0.157 $\pm$    0.061 &     2.49 $\pm$     0.92 \\
   46.20 &    0.141 $\pm$    0.069 &     2.61 $\pm$     1.03 \\
   50.49 &    0.192 $\pm$    0.083 &     3.33 $\pm$     1.25 \\
   54.78 &    0.198 $\pm$    0.105 &     2.60 $\pm$     1.59 \\
   59.07 &    0.223 $\pm$    0.129 &     3.33 $\pm$     1.99 \\
\hline
\end{tabular}
\end{table}

\begin{table}
\caption{  NGC~6221}
\begin{tabular}{lcc}  \hline
Radius    & Mg$_2$ & Fe5270 \\ 
(arcsec)  & (mag) & ($\AA$) \\
\hline
    0.0  &    0.168 $\pm$    0.033 &     1.95 $\pm$     0.31 \\
    3.3  &    0.164 $\pm$    0.034 &     1.70 $\pm$     0.32 \\
    7.59 &    0.187 $\pm$    0.035 &     1.15 $\pm$     0.32 \\
   11.88 &    0.183 $\pm$    0.037 &     1.61 $\pm$     0.36 \\
   16.17 &    0.176 $\pm$    0.039 &     1.59 $\pm$     0.38 \\
   20.46 &    0.176 $\pm$    0.041 &     1.30 $\pm$     0.40 \\
   24.75 &    0.149 $\pm$    0.043 &     1.46 $\pm$     0.46 \\
   29.04 &    0.141 $\pm$    0.045 &     1.35 $\pm$     0.49 \\
   33.33 &    0.178 $\pm$    0.048 &     0.98 $\pm$     0.51 \\
   37.62 &    0.166 $\pm$    0.051 &     1.32 $\pm$     0.56 \\
   41.91 &    0.159 $\pm$    0.053 &     0.90 $\pm$     0.57 \\
   46.20 &    0.166 $\pm$    0.054 &     0.80 $\pm$     0.62 \\
   50.49 &    0.173 $\pm$    0.057 &     0.54 $\pm$     0.64 \\
   54.78 &    0.169 $\pm$    0.057 &     1.19 $\pm$     0.64 \\
   59.07 &    0.162 $\pm$    0.063 &     1.09 $\pm$     0.69 \\
\hline
\end{tabular}
\end{table}

\begin{table}
\caption{  NGC~6753}
\begin{tabular}{lcc}  \hline
Radius    & Mg$_2$ & Fe5270 \\ 
(arcsec)  & (mag) & ($\AA$) \\
\hline
    0.0  &    0.396 $\pm$    0.032 &     2.58 $\pm$     0.37 \\
    3.3  &    0.320 $\pm$    0.032 &     2.82 $\pm$     0.38 \\
    7.59 &    0.265 $\pm$    0.032 &     2.59 $\pm$     0.38 \\
   11.88 &    0.240 $\pm$    0.033 &     2.91 $\pm$     0.41 \\
   16.17 &    0.243 $\pm$    0.033 &     2.62 $\pm$     0.42 \\
   20.46 &    0.235 $\pm$    0.035 &     2.28 $\pm$     0.45 \\
   24.75 &    0.236 $\pm$    0.036 &     2.36 $\pm$     0.49 \\
   29.04 &    0.233 $\pm$    0.037 &     2.05 $\pm$     0.53 \\
   33.33 &    0.228 $\pm$    0.040 &     2.27 $\pm$     0.63 \\
   37.62 &    0.238 $\pm$    0.045 &     2.40 $\pm$     0.76 \\
   41.91 &    0.266 $\pm$    0.051 &     2.37 $\pm$     0.92 \\
   46.20 &    0.271 $\pm$    0.055 &     2.14 $\pm$     1.05 \\
   50.49 &    0.250 $\pm$    0.058 &     2.57 $\pm$     1.11 \\
   54.78 &    0.213 $\pm$    0.061 &     1.87 $\pm$     1.22 \\
   59.07 &    0.218 $\pm$    0.068 &     1.99 $\pm$     1.39 \\
\hline
\end{tabular}
\end{table}

\begin{table}
\caption{  NGC~6814}
\begin{tabular}{lcc}  \hline
Radius    & Mg$_2$ & Fe5270 \\ 
(arcsec)  & (mag) & ($\AA$) \\
\hline
    0.0  &    0.226 $\pm$    0.031 &     1.00 $\pm$     0.27 \\
    3.3  &    0.283 $\pm$    0.032 &     2.01 $\pm$     0.33 \\
    7.59 &    0.273 $\pm$    0.034 &     2.11 $\pm$     0.37 \\
   11.88 &    0.239 $\pm$    0.036 &     1.83 $\pm$     0.42 \\
   16.17 &    0.254 $\pm$    0.039 &     2.37 $\pm$     0.49 \\
   20.46 &    0.241 $\pm$    0.039 &     2.13 $\pm$     0.50 \\
   24.75 &    0.236 $\pm$    0.040 &     1.70 $\pm$     0.52 \\
   29.04 &    0.208 $\pm$    0.043 &     1.51 $\pm$     0.59 \\
   33.33 &    0.233 $\pm$    0.046 &     1.52 $\pm$     0.65 \\
   37.62 &    0.241 $\pm$    0.052 &     0.81 $\pm$     0.79 \\
   41.91 &    0.248 $\pm$    0.062 &     1.69 $\pm$     0.98 \\
   46.20 &    0.282 $\pm$    0.071 &     0.87 $\pm$     1.19 \\
   50.49 &    0.207 $\pm$    0.080 &     1.43 $\pm$     1.35 \\
   54.78 &    0.249 $\pm$    0.089 &     1.34 $\pm$     1.48 \\
   59.07 &    0.208 $\pm$    0.098 &     1.60 $\pm$     1.68 \\
\hline
\end{tabular}
\end{table}

\begin{table}
\caption{  NGC~6935}
{\tiny
\begin{tabular}{lcc}  \hline
Radius    & Mg$_2$ & Fe5270 \\ 
(arcsec)  & (mag) & ($\AA$) \\
\hline
    0.0  &    0.294 $\pm$    0.031 &     2.42 $\pm$     0.36 \\
    2.31 &    0.283 $\pm$    0.032 &     1.83 $\pm$     0.35 \\
    4.62 &    0.297 $\pm$    0.032 &     2.37 $\pm$     0.38 \\
    6.93 &    0.302 $\pm$    0.033 &     2.60 $\pm$     0.40 \\
    9.24 &    0.322 $\pm$    0.033 &     2.54 $\pm$     0.41 \\
   11.55 &    0.311 $\pm$    0.034 &     2.62 $\pm$     0.42 \\
   13.86 &    0.298 $\pm$    0.034 &     2.69 $\pm$     0.43 \\
   16.17 &    0.269 $\pm$    0.035 &     2.15 $\pm$     0.42 \\
   18.48 &    0.250 $\pm$    0.035 &     2.50 $\pm$     0.44 \\
   20.79 &    0.246 $\pm$    0.036 &     2.19 $\pm$     0.44 \\
   23.10 &    0.268 $\pm$    0.038 &     1.96 $\pm$     0.47 \\
   25.41 &    0.260 $\pm$    0.041 &     2.25 $\pm$     0.53 \\
   27.72 &    0.268 $\pm$    0.045 &     2.28 $\pm$     0.57 \\
   30.03 &    0.264 $\pm$    0.049 &     1.89 $\pm$     0.63 \\
   32.34 &    0.241 $\pm$    0.052 &     1.76 $\pm$     0.69 \\
   34.65 &    0.258 $\pm$    0.058 &     1.67 $\pm$     0.77 \\
   36.96 &    0.288 $\pm$    0.065 &     2.01 $\pm$     0.86 \\
   39.27 &    0.271 $\pm$    0.073 &     2.32 $\pm$     0.97 \\
   41.58 &    0.296 $\pm$    0.087 &     1.66 $\pm$     1.14 \\
   43.89 &    0.333 $\pm$    0.105 &     1.71 $\pm$     1.41 \\
   46.20 &    0.285 $\pm$    0.120 &     2.26 $\pm$     1.60 \\
   48.51 &    0.346 $\pm$    0.141 &     1.77 $\pm$     1.85 \\
   50.82 &    0.370 $\pm$    0.160 &     0.51 $\pm$     2.13 \\
   53.13 &    0.304 $\pm$    0.179 &     1.10 $\pm$     2.38 \\
   55.44 &    0.340 $\pm$    0.208 &     1.88 $\pm$     2.73 \\
   57.75 &    0.361 $\pm$    0.241 &     2.32 $\pm$     2.84 \\
\hline
\end{tabular}
}
\end{table}

\begin{table}
\caption{  NGC~7213}
\begin{tabular}{lcc}  \hline
Radius    & Mg$_2$ & Fe5270 \\ 
(arcsec)  & (mag) & ($\AA$) \\
\hline
    0.0  &    0.231 $\pm$    0.031 &     1.67 $\pm$     0.30 \\
    3.3  &    0.232 $\pm$    0.031 &     1.71 $\pm$     0.30 \\
    7.59 &    0.202 $\pm$    0.031 &     1.64 $\pm$     0.31 \\
   11.88 &    0.177 $\pm$    0.031 &     1.37 $\pm$     0.30 \\
   16.17 &    0.175 $\pm$    0.032 &     1.48 $\pm$     0.34 \\
   20.46 &    0.187 $\pm$    0.034 &     1.61 $\pm$     0.39 \\
   24.75 &    0.187 $\pm$    0.036 &     1.33 $\pm$     0.46 \\
   29.04 &    0.200 $\pm$    0.040 &     1.51 $\pm$     0.57 \\
   33.33 &    0.210 $\pm$    0.044 &     1.78 $\pm$     0.67 \\
   37.62 &    0.204 $\pm$    0.049 &     2.00 $\pm$     0.78 \\
   41.91 &    0.235 $\pm$    0.055 &     1.69 $\pm$     0.91 \\
   46.20 &    0.243 $\pm$    0.061 &     2.20 $\pm$     1.05 \\
   50.49 &    0.246 $\pm$    0.070 &     2.39 $\pm$     1.20 \\
   54.78 &    0.278 $\pm$    0.079 &     2.82 $\pm$     1.37 \\
   59.07 &    0.307 $\pm$    0.090 &     2.95 $\pm$     1.57 \\
\hline
\end{tabular}
\end{table}

\begin{table}
\caption{  NGC~7412}
\begin{tabular}{lcc}  \hline
Radius    & Mg$_2$ & Fe5270 \\ 
(arcsec)  & (mag) & ($\AA$) \\
\hline
    0.0  &    0.158 $\pm$    0.034 &     1.99 $\pm$     0.35 \\
    3.3  &    0.154 $\pm$    0.037 &     2.03 $\pm$     0.40 \\
    7.59 &    0.145 $\pm$    0.043 &     1.93 $\pm$     0.51 \\
   11.88 &    0.149 $\pm$    0.048 &     1.96 $\pm$     0.61 \\
   16.17 &    0.136 $\pm$    0.054 &     2.04 $\pm$     0.70 \\
   20.46 &    0.152 $\pm$    0.060 &     1.90 $\pm$     0.80 \\
   24.75 &    0.136 $\pm$    0.063 &     2.00 $\pm$     0.86 \\
   29.04 &    0.147 $\pm$    0.066 &     1.80 $\pm$     0.90 \\
   33.33 &    0.163 $\pm$    0.067 &     1.70 $\pm$     0.91 \\
   37.62 &    0.138 $\pm$    0.071 &     1.55 $\pm$     0.96 \\
   41.91 &    0.140 $\pm$    0.082 &     2.23 $\pm$     1.12 \\
   46.20 &    0.118 $\pm$    0.100 &     1.46 $\pm$     1.43 \\
   50.49 &    0.133 $\pm$    0.128 &     1.66 $\pm$     1.84 \\
   54.78 &    0.157 $\pm$    0.173 &     1.87 $\pm$     2.41 \\
   59.07 &    0.137 $\pm$    0.230 &     1.06 $\pm$     3.28 \\
\hline
\end{tabular}
\end{table}

\begin{table}
\caption{  NGC~7637}
\begin{tabular}{lcc}  \hline
Radius    & Mg$_2$ & Fe5270 \\ 
(arcsec)  & (mag) & ($\AA$) \\
\hline
    0.0  &    0.237 $\pm$    0.031 &     2.20 $\pm$     0.33 \\
    3.3  &    0.253 $\pm$    0.032 &     2.94 $\pm$     0.37 \\
    7.59 &    0.233 $\pm$    0.033 &     2.65 $\pm$     0.39 \\
   11.88 &    0.193 $\pm$    0.034 &     2.50 $\pm$     0.42 \\
   16.17 &    0.196 $\pm$    0.036 &     1.93 $\pm$     0.43 \\
   20.46 &    0.219 $\pm$    0.038 &     1.79 $\pm$     0.49 \\
   24.75 &    0.210 $\pm$    0.041 &     2.52 $\pm$     0.56 \\
   29.04 &    0.187 $\pm$    0.044 &     2.41 $\pm$     0.63 \\
   33.33 &    0.178 $\pm$    0.051 &     2.25 $\pm$     0.80 \\
   37.62 &    0.232 $\pm$    0.064 &     3.39 $\pm$     1.00 \\
   41.91 &    0.225 $\pm$    0.071 &     2.43 $\pm$     1.18 \\
   46.20 &    0.200 $\pm$    0.090 &     2.73 $\pm$     1.52 \\
   50.49 &    0.205 $\pm$    0.101 &     3.47 $\pm$     1.75 \\
   54.78 &    0.209 $\pm$    0.112 &     2.91 $\pm$     1.95 \\
   59.07 &    0.203 $\pm$    0.141 &     4.42 $\pm$     2.38 \\
\hline
\end{tabular}
\end{table}

\end{document}